\documentclass[journal]{IEEEtran}

\usepackage{amsmath,amssymb,amsfonts}
\usepackage{algorithmic}
\usepackage{graphicx}
\usepackage{textcomp}
\usepackage{xcolor}
\usepackage{multirow} % Needed for multirow cells in the table
\usepackage{booktabs}
\usepackage{booktabs}
\usepackage{xcolor}
\usepackage{siunitx}
\usepackage{acronym}
\usepackage{makecell}
\usepackage{cuted}
\usepackage{booktabs}
\usepackage{dblfloatfix}
\usepackage{lipsum}
\usepackage{dblfloatfix}
\usepackage{comment}
\usepackage{cite}

\acrodef{ode}[ODE]{Ordinary Differential Equation}
\acrodef{dae}[DAE]{Differential Algebraic Equation}
\acrodef{rms}[RMS]{Root Mean Square}
\acrodef{emt}[EMT]{Electromagnetic Transient}
\acrodef{tf}[TF]{Transfer Function}
\acrodef{vsc}[VSC]{Voltage Source Converter}
\acrodef{gfm}[GFM]{Grid-Forming}
\acrodef{gfl}[GFL]{Grid-Following}
\acrodef{sg}[SG]{Synchronous Generator}
\acrodef{stamp}[STAMP]{Small-Signal Toolbox for Analysis of Modern Power Systems}
\acrodef{tds}[TDS]{Time Domain Simulation}
\acrodef{ibr}[IBR]{Inverter-Based Resource}
\acrodef{siso}[SISO]{Single-Input-Single-Output}
\acrodef{mimo}[MIMO]{Multiple-Input-Multiple-Output}
\acrodef{fsi}[FSI]{Frequency Sensitivity Indicator}
\acrodef{vsi}[VSI]{Voltage Sensitivity Indicator}
\acrodef{svd}[SVD]{Singular Value Decomposition}

\begin{document}
%

%\title{Power System Forced Oscillation Detection and Propagation Analysis Based on Voltage and Frequency Dynamics}
\title{Quantifying Power to Voltage and Frequency Dynamics for
Oscillation 
%Detection and 
Propagation Assessment}
% \title{Forced Oscillation Detection and Propagation Analysis Based on Apparent Power to Voltage and Frequency Transfer Functions}

%
%
\author{Onur Alican,~\IEEEmembership{Student Member,~IEEE,}
        Dionysios Moutevelis,~\IEEEmembership{Member,~IEEE,}
        Marc Cheah-Mañe,~\IEEEmembership{Senior Member,~IEEE,}
        Oriol Gomis-Bellmunt,~\IEEEmembership{Fellow,~IEEE,} and
        Eduardo Prieto-Araujo,~\IEEEmembership{Senior Member,~IEEE}
   }

\maketitle

\begin{abstract}
%
%Modern electrical power systems are undergoing a dual transition: a structural expansion driven by increasing demand and a dynamic evolution caused by the integration of fast-acting Inverter-Based Resources (\acp{ibr}).
%
The integration of Inverter-Based Resources (IBRs) into power systems introduces multi-timescale dynamics 
%and increase the need for localized, bus-level dynamic awareness, as oscillations originating in one region
and oscillations which may propagate to distant areas and endanger the safe system operation.
%ead to harmful impacts} on the power system components in the worst case, blackouts.
%
%Forced oscillations pose a particularly significant challenge under these conditions
%
These oscillations pose a significant challenge as their source, frequency, and propagation pathways are often challenging to identify in large interconnected systems, comprising numerous synchronous machines and IBRs.
%
% Additionally, the coupling between voltage and frequency variables further impedes the isolation of the oscillation root cause, especially when non-uniform theoretical approaches are used for the analysis of each case.
%
For the above reasons, analytical tools that identify the sensitivity of the system to oscillations within a large frequency spectrum, affecting both voltage and frequency variables across various network locations, are of interest.
This paper addresses this topic by introducing a frequency-domain framework based on linear analysis, which characterizes the sensitivity to oscillations of each network bus voltage and of each generation unit frequency.
%
%sensitivity dynamics through the grid buses, and proposes
%
Within this framework, two quantitative indicators are proposed, namely the Frequency Sensitivity Index (FSI) and Voltage Sensitivity Index (VSI).
These indexes are derived analytically from the transfer functions which relate the active and reactive power injections to each bus with the voltage and frequency variables across the network, derived from the linear Electromagnetic Transient (EMT) power system model.
%outputs) and are obtained via Electromagnetic Transient (EMT) modeling.
%
%These indices enable systematic frequency-domain screening to understand the impact that active and reactive power injections at any bus have on the bus voltages and generator frequencies.
%
The proposed indices quantify the sensitivity of the system to oscillations of different type and frequency, providing insights for both oscillation detection and propagation analysis.
The methodology is applied to a case study based on the modified IEEE 68-bus benchmark system under partial and full IBR penetration, while its accuracy is validated through EMT time-domain simulations using linear and nonlinear models developed in Matlab/Simulink environment.
%
%The proposed methodology \textcolor{black}{is aimed to be used by} Transmission System Operators (TSOs) for system level planning and security assessment.
\end{abstract}


\IEEEpeerreviewmaketitle

\section{Introduction}
%

%
% \textcolor{red}{General comment on the motivation subsection. In general it is very long, it should be more to the point. The main narrative we try to push is:
% IBR integration-->>problems-->forced oscillations-->new tools are necessary, if analytical even better-->we propose a new analytical tool that is amazing, but we do not describe very much. This will be later done in the contribution subsection}
%
Electrical power systems are undergoing a dual transition, a structural expansion to meet increasing demand and a shift in their dynamic operation, driven by the integration of \acp{ibr}  which operate at broader timescales than conventional generation.
These changes necessitate 
%
%frequency-domain analysis to
%
novel analytic tools that represent the system dynamic behavior across multiple timescales~\cite{low_inertia}. At the same time, network expansion increases the need for localized, bus-level information that can provide insights about the dynamic operation of the system that consider its topological characteristics~\cite{expandinggrid}.

Within this evolving environment, oscillations present a particularly challenging threat, and are classified based on both their type and spatial characteristics~\cite{cheng2022real}. 
Regarding their type, oscillations can be classified between natural and forced oscillations~\cite{ghorbaniparvar2017survey}.
Natural oscillations arise from the system’s inherent dynamics (its modes and eigenvalues), whereas forced oscillations are driven by external periodic disturbances that impose an oscillation frequency on the system, in addition to its natural modes~\cite{forced_equip}.
% Forced oscillations may rise from malfunctioning equipment, incorrect operating conditions, or periodic load variations that create perpetual periodic disturbances in to the power system.
%
Regarding their spatial distribution across the network, oscillations can either remain local to specific areas of the power system or spread through the electrical network~\cite{mateu2025power,moutevelis2024modal}.
%
%The latter category poses significant risks, including damage to equipment, unintended triggering of protection or control actions, reduced transmission line capacity, component failures, and, in extreme cases, a complete system blackout.
%
%
% Oscillations \textcolor{black}{usually remain in local areas}; however, they might be spread through the electrical network.
%
%In both localized and widespread scenarios, cases that threaten the vital variables, such as voltage and frequency, and are dangerous to the power system appear~\cite{_2017_reliability}.
%
In both local and system-level propagation scenarios, oscillations may lead to critical failures of essential power system components, posing a threat to the safe system operation~\cite{_2017_reliability}.
%
% , as identifying these oscillations remains challenging due to the difficulty in determining their source location, frequency, and amplitude in large interconnected systems~\cite{_2017_reliability}}.
%
%
For the above reasons, observing oscillations of system variables, such as frequency and voltage, becomes increasingly critical, motivating the development of analytical approaches that quantify the sensitivity of the system to such oscillations.
%
%and assess their dynamic response, motivating the analysis of power to voltage and frequency dynamics (also rephrased as $PQ\text{-}V\omega$ transfer function in this paper) of generation units and bus voltages in the frequency domain.
%
%

%
% \textcolor{red}{General comments on the literature review:
% \begin{enumerate}
%     % \item The literature review can get a bit stronger, 10 references are not a lot and the topic of oscillation analysis is very well studied. E.g. I cite here some references that I found with a 5 minute search~\cite{ye2016analysis,ghorbaniparvar2017survey,follum2015detection,wang2017location,wu2019distributed,zhi2020analysis,huang2020analytical,jiang2000adaptive,ma2010application,moutevelis2024modal}.
%     \item The flow of the review feels a bit reverse than it should. You start very specific with the PQ-wV tools that have been proposed and then you go general for other oscillation identification methods. I think it should be the other way around, from general to specific
%     \item You can address the 2nd point by grouping the references to paragraphs and explaining the general methodology (e..g, data based from PMUs, energy based etc.) and why it may have problems. E.g., data based methods require data that we may not have, thus modelling based approaches are useful. No need to "attack" each reference directly, only the ones that are very similar to our work, i.e., they use the same method for the same problem.
% \end{enumerate}}
% %
%
In broad terms, the approaches to study oscillations in power systems can be categorized between analytic and measurement-based ones.
Traditionally, analytic studies of oscillation propagation in power systems has focused on \ac{sg}-dominated networks, as in \cite{analysis_of_dist_pro,analytical_propagation,samyaktamrakar_2018_propagation}. These studies typically rely on simplified models, assuming homogeneous \ac{sg} parameters, neglecting ohmic losses, and considering constant voltage magnitudes across the network.
%
%Consequently, their accuracy depends strongly on the underlying network topology.
%
With the increasing integration of \acp{ibr}, more advanced techniques that account for the increased coupling between voltage and frequency variables have been presented in the literature.
For instance,~\cite{moutevelis2024modal} analyzes oscillation propagation using participation factors of complex frequency variables in both conventional and \ac{ibr}-rich systems. However, this approach is based on algebraic network equations, i.e., the \ac{rms} modeling framework, which provides limited insight for high-frequency oscillation phenomena.

The widespread deployment of Phasor Measurement Units (PMUs) has enabled advanced measurement-based oscillation analysis~\cite{phadke2018phasor}. Research has progressed from oscillation source identification using energy-flow- and mode-shape-based methods~\cite{def,wang_2016_location}, to model-based approaches integrating PMU data with dynamic models~\cite{agrawal_2017}, and more recently to data-driven techniques such as machine learning, dynamic mode decomposition, and sparse system identification~\cite{syn_phasor_data,ML}. However, these methods primarily identify oscillation sources rather than their propagation throughout the network.

Recently, the transfer function that relates the active and reactive power injections to the voltage and frequency variables of the grid, also termed as the the $PQ\text{-}V\omega$ Multiple-Input Multiple-Output (MIMO) transfer function, has been shown to be a promising theoretical tool for the study of power system dynamics and stability, and has been used at both the power system and component levels~\cite{wp_unified}.
At the device level, references~\cite{pw_gen_pov,yang2025impact} employ a coupled voltage and frequency theoretical framework to assess and improve the control performance of virtual synchronous machines. However, these works are limited to device-level applications and do not extend to system-level analysis.

Moving to small-scale system studies and to the interaction of a limited number of converters,~\cite{pw_gen_networksmall} uses a similar modeling approach to investigate oscillations in multi-converter systems connected to an ideal grid.
In the context of multi-mode converter control interactions,~\cite{Pw_mppt} examines a new type of forced oscillation involving \ac{gfl} and \ac{gfm} converters through the $P\text{-}\omega$ transfer function.
However, the presented analysis is restricted to specific converter control structures
%configurations that involve converters equipped with specific mixed control structures 
and lacks system-level applicability.

Regarding system-oriented studies,~\cite{wp_Onur} analyzes the $P\text{-}\omega$ relationship in a general power system by aggregating the combined dynamic behaviour of all system generator rotor speeds to a scalar metric and examines two options for this metric, namely 
%treating frequency as a global variable, 
%and compares two frequency-domain metrics:
the maximum singular value $\sigma_{\max}$ and the Center of Inertia.
However, the scope of the work was focused to active power-to-frequency relationships, neglecting reactive power-to-voltage, as well as the frequency/voltage coupling.
This paper introduces two novel quantitative indices, namely the \ac{fsi} and the \ac{vsi}, aimed at identifying oscillations of both voltage and frequency variables in power systems, quantifying the system impact to these oscillations across a frequency range and tracking their propagation across the network.
%
% \textcolor{red}{Maybe we can call them Frequency/Voltage \underline{sensitivity} Indicators? Maybe it is more fitting and avoids the whole Sensitivity discussion. Also, it keeps the same acronym :-)}
%to characterize the $PQ\text{-}V\omega$ transfer function and support a systematic assessment of these dynamic interactions.
%
These frequency-domain metrics are extracted from the MIMO $PQ-V\omega$ transfer function, linking the active and reactive power injections to all system buses with the bus voltages and generator rotor speeds across the system.
%
%\textcolor{red}{We should add a sentence here on whether the oscillations that are adressed are natural, forced or both. This has to be cleared up with everyone.}
%
The proposed indices provide a comprehensive framework for offline system studies, capable of assessing grid vulnerability to both internal natural modes and externally imposed periodic forced oscillations.
%, as well as the relation between the two 
%
%The proposed analysis enables the diagnosis of critical system variables and the comprehensive analysis of the system oscillatory behaviour.
%ion propagation across the network.
%and computed for all the buses of the EMT power system model.
%
% The main contributions of this work are summarized as follows:
% %
% \begin{itemize}
%         \item Analytical extraction of $PQ\!-\!V\omega$ transfer functions, computation of Frequency and Voltage Sensitivity Indicators for all buses in the network including generation and passive buses, and establishing their analytical link to system eigenproperties.
%         \item Proposing two novel quantitative frequency domain voltage and frequency Sensitivity indicators and applying them for natural oscillating source identification and oscillation propagation
% \end{itemize}
%
%
%
% \textcolor{red}{I think the bullets can be reintroduced, but with a better explanation of the contents of Section III, for example:
% \begin{itemize}
%     \item Using the global transfer function (this not a novelty of the work per se), we introduce the new metrics
%     \item we introduce the global metrics
%     \item we reveal their relations with the eigenvectors
% \end{itemize}
The main contributions of this work are summarized as follows:
\begin{itemize}
\item Starting from the $PQ\text{-}V\omega$ MIMO transfer function of the complete power system, 
%two novel frequency-domain indices are introduced, namely
the Frequency Sensitivity Indicator (FSI) and the Voltage Sensitivity Indicator (VSI) are introduced.
These metrics quantify the impact of oscillatory dynamics on the frequency- and voltage-related variables over a wide frequency range.
\item Global and local formulations of the proposed metrics are developed, enabling both general and detailed assessment of the system oscillatory behaviour.
%, enabling the identification of dominant oscillatory modes, the assessment of their spatial propagation across the network, and the diagnosis of the most critical buses and generators contributing to the observed dynamics.
\item The relationship between the proposed metrics and the modal properties of the system is analytically presented through their connection with the system eigenstructure.
%, showing how peaks in the FSI and VSI are associated with natural oscillatory modes and their corresponding eigenvectors.
\end{itemize}

Contrary to other measurement-based oscillation studies found in the literature, the objective of this work is not to determine which system component is the source, but rather to assess how oscillatory phenomena propagate throughout the network.
%and which buses are dynamically most exposed to them.
%
Furthermore, unlike purely measurement-based approaches, the proposed methodology is model-based and exploits the dynamic representation of the system to quantify the propagation of oscillatory interactions across the network. As a result, the proposed framework provides information that is complementary to oscillation source identification methods, enabling the identification of dynamically vulnerable locations and the characterization of oscillation transmission paths.
%in addition to the source itself.
%
The applicability of the proposed metrics %for natural oscillation source detection and propagation analysis 
is demonstrated in two case studies based on modified versions of the well-known IEEE 68-bus benchmark systems under partial and full \ac{ibr} penetration.
%
%representing realistic and future power system scenarios, and
%
The results of the analysis are validated through time-domain EMT simulations, using both linear and nonlinear models, developed in Matlab/Simulink environment.
%
% The above will make the reader be aware of the contents of the main section of the paper which is the core contribution, but without repeating the same info several times.
% }
%
% \section{Methodology}
% %
% In this section, the methodology for obtaining the $PQ-V\omega$ transfer function analytically from generic power system state-space models is first presented.
% %
% Then, the derivation of the two proposed metrics, \ac{fsi} and \ac{vsi}, from the $PQ-V\omega$ transfer function will be demonstrated, which is the main contribution of this work.
%
\vspace{-0.2cm}
\section{$PQ-V\omega$ transfer function}
%\vspace{-0.2cm}
\subsection{Small-Signal Preliminaries}
The linear model of a power system, resulting after linearizing the original, nonlinear \ac{ode} system around a specific operating point, can be written in the following state-space form~\cite{ogata2010modern}:
\begin{equation}
\label{eq:linear_ode}
\Delta \dot{\boldsymbol{x}} = \boldsymbol{A}\Delta \boldsymbol{x} + \boldsymbol{B}\Delta \boldsymbol{u}, \qquad
\Delta \boldsymbol{y} = \boldsymbol{C}\Delta \boldsymbol{x} + \boldsymbol{D}\Delta \boldsymbol{u}.
\end{equation}
where $\boldsymbol{x}$, $\boldsymbol{u}$ and $\boldsymbol{y}$ are the state variable, input and output vectors, respectively, while $\boldsymbol{A}$, $\boldsymbol{B}$, $\boldsymbol{C}$ and $\boldsymbol{D}$ are the state, input, output and feedforward matrices, respectively.
$\Delta$ denotes a small perturbation of the respective variable.
The necessary matrices in~\eqref{eq:linear_ode} can be calculated at device level and then used to synthesize the complete system linear model~\cite{collados2024stability,wang2017small}.
When the grid and other relevant dynamics in the electromagnetic scale are included \textcolor{black}{as differential equations in}~\eqref{eq:linear_ode}, this representation is often referred to as the EMT state-space model of the network~\cite{lacerda2022phasor,cheah2026new}, which is the approach selected in this work.
%
%This approach allows accurate modelling of oscillatory phenomena in the power system up to frequencies that are significantly larger compared to the conventional, phasor-based modelling.
%
%\subsection{$P$-$\omega$ Transfer Function Calculation}
%
% This subsection explains the methodology to extract $P$-$\omega$ dynamics from the generic model of~\eqref{eq:linear_ode}.
% %
% The objective is to evaluate how an active power injection $P$ at each bus would affect the frequency $\omega$ of each generation unit (\acp{sg} and \acp{ibr} controlled in GFM, and GFL modes). 
% %
% %This reveals how power injections at different buses and frequencies affect generator rotor speeds. 
%

\color{black}
In the proposed state-space framework, active and reactive power injections $(P, Q)$ to each bus are treated as system inputs and represent external disturbances acting on the power system, while bus voltage magnitudes and generator rotor speeds ($V,\omega$, respectively) are selected as outputs.
These outputs are selected as they are observable variables that fully capture the response with regards to voltage and frequency oscillatory disturbances across the whole network.
\vspace{-0.35cm}
\subsection{Transfer Function Derivation}

The transformation between active and reactive power and current variables is presented as follows, starting from the expression of active and reactive power in a $qd$-reference frame:
\begin{align}
\label{eq.pq_injections}
    P 
    =\frac{3}{2}(v_{q}i_{q}+v_{d}i_{d})
    ,
    \;
    Q 
    =
    \frac{3}{2}(v_{q} i_{d} - v_{d} i_{q})
    ,
\end{align}
%
%where the bus number is denoted as a subscript.
%
where variables $i_{qd}$ and $v_{qd}$ are the current injection and bus voltage $qd$-components for every bus, respectively. 
%
%By setting $v=[v_q \; v_d]^\top$, $v=[v_q \; v_d]^\top$
%
By organizing in matrix form and linearizing, \eqref{eq.pq_injections} becomes:
% %
% \begin{equation}
% \label{eq:pq_injections_matrix}
% \begin{aligned}
% \begin{bmatrix}
% P\\
% Q
% \end{bmatrix}
% &=
% \frac{3}{2}
% \begin{bmatrix}
% v_q & v_d\\
% -v_d & v_q
% \end{bmatrix}
% \begin{bmatrix}
% i_q\\
% i_d
% \end{bmatrix}
% .
% \end{aligned}
% \end{equation}
% %
% By linearizing~\eqref{eq:pq_injections_matrix}, the following expression is obtained:
%
\begin{equation}
\label{eq:pq_injections_matrix_linear}
\begin{bmatrix}
\Delta P\\
\Delta Q
\end{bmatrix}
=
\frac{3}{2}
\left(
\begin{bmatrix}
v_{q0} & v_{d0}\\
-v_{d0} & v_{q0}
\end{bmatrix}
\begin{bmatrix}
\Delta i_q\\
\Delta i_d
\end{bmatrix}
+
\begin{bmatrix}
\Delta v_q & \Delta v_d\\
- \Delta v_d & \Delta v_q
\end{bmatrix}
\begin{bmatrix}
i_{q0}\\
i_{d0}
\end{bmatrix}
\right),
\end{equation}
where index $0$ signifies the linearization point. Solving for the current, one gets the current injections at each bus as a linear combination of active and reactive power disturbances and the voltage components of said bus:
\begin{equation}
\small
\label{eq:deltai_final}
\begin{aligned}
\begin{bmatrix}
\Delta i_{q}\\
\Delta i_{d}
\end{bmatrix}&
=
\frac{1}{v_{q0}^2+v_{d0}^2}
\begin{bmatrix}
v_{q0} & -v_{d0}\\
v_{d0} & v_{q0}
\end{bmatrix}
\cdot
\\
&
\left(
\frac{2}{3}
\begin{bmatrix}
\Delta P
\\
\Delta Q
\end{bmatrix}
-
\begin{bmatrix}
\Delta v_q & \Delta v_d\\
-\Delta v_d & \Delta v_q
\end{bmatrix}
\begin{bmatrix}
i_{q0}\\
i_{d0}
\end{bmatrix}
\right).
\end{aligned}
\end{equation}

\begin{figure}[t!]
    \centering
    \includegraphics[width=0.9\columnwidth]{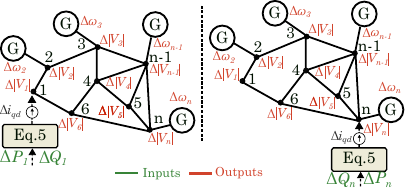}
    \vspace{-0.35cm}
   % \captionsetup{font=footnotesize}
    \caption{$P$ and $Q$ disturbance to each bus and obtain $\omega$ of generation units and voltage magnitude of each bus}
    \label{fig:p_w_fig}
\end{figure}

Fig.~\ref{fig:p_w_fig} illustrates $\Delta P$ and $\Delta Q$ inputs as virtual current injections and the outputs.
This is achieved by means of combining~\eqref{eq:deltai_final} and virtual shunt current sources in each bus, similar to the process followed for frequency scanning by means of sinusoidal current injections in impedance studies~\cite{zhu2026hybrid}. 
Once the $PQ$-$i_{qd}$ relationship is obtained, the active and reactive power injections as a disturbance at every system bus are selected as inputs to the full system state space model of~\eqref{eq:linear_ode}: $\boldsymbol{u}=[\Delta P_1\dots\Delta P_n \; \Delta Q_1 \dots \Delta Q_n]^\top$, where $n$ is the number of network buses.
At the same time, the rotor speeds of the generation units and the voltage magnitudes ($|\cdot|$ denotes the magnitude of the corresponding Park vector) of each bus are selected as outputs: $\boldsymbol{y}=[\Delta \omega_1\dots \Delta \omega_m  \; \Delta|V_1| \dots \Delta |V_n|]^\top$, where $m$ is the number of generators in the network.
It is noted that for \ac{ibr}-based generators, which do not inherently possess a rotor, the internal frequencies generated by the converter controller are used~\cite{moutevelis2023taxonomy}.
By substituting $\boldsymbol{u}$ and $\boldsymbol{y}$ in~\eqref{eq:linear_ode} and converting from a time-domain to a transfer function representation: $\boldsymbol{G}(s)=\boldsymbol{C}(s\boldsymbol{I}-\boldsymbol{A})^{-1}\boldsymbol{B}+\boldsymbol{D}$, the $PQ-V\omega$ transfer functions that relates all the active and reactive power injections in the system to all generator frequencies and bus voltage magnitude is obtained:
\vspace{-0.4cm}
\begin{equation}
\footnotesize
\setlength{\arraycolsep}{1.5pt}
\renewcommand{\arraystretch}{0.9}
\label{eq.pq_vw_matrix}
\boldsymbol{y}
=
\begin{bmatrix}
\underbrace{
\begin{array}{@{}c@{\hspace{2pt}}c@{\hspace{2pt}}c@{}}
G_{\omega_1,P_1}(s) & \cdots & G_{\omega_1,P_n}(s)\\
\vdots & \ddots & \vdots\\
G_{\omega_m,P_1}(s) & \cdots & G_{\omega_m,P_n}(s)
\end{array}
}_{\boldsymbol{G}_{\omega,P}(s)}
&
\underbrace{
\begin{array}{@{}c@{\hspace{2pt}}c@{\hspace{2pt}}c@{}}
G_{\omega_1,Q_1}(s) & \cdots & G_{\omega_1,Q_n}(s)\\
\vdots & \ddots & \vdots\\
G_{\omega_m,Q_1}(s) & \cdots & G_{\omega_m,Q_n}(s)
\end{array}
}_{\boldsymbol{G}_{\omega,Q}(s)}
\\[2pt]
\underbrace{
\begin{array}{@{}c@{\hspace{2pt}}c@{\hspace{2pt}}c@{}}
G_{|V_1|,P_1}(s) & \cdots & G_{|V_1|,P_n}(s)\\
\vdots & \ddots & \vdots\\
G_{|V_n|,P_1}(s) & \cdots & G_{|V_n|,P_n}(s)
\end{array}
}_{\boldsymbol{G}_{|V|,P}(s)}
&
\underbrace{
\begin{array}{@{}c@{\hspace{2pt}}c@{\hspace{2pt}}c@{}}
G_{|V_1|,Q_1}(s) & \cdots & G_{|V_1|,Q_n}(s)\\
\vdots & \ddots & \vdots\\
G_{|V_n|,Q_1}(s) & \cdots & G_{|V_n|,Q_n}(s)
\end{array}
}_{\boldsymbol{G}_{|V|,Q}(s)}
\end{bmatrix}
\boldsymbol{u}.
\end{equation}
\vspace{-0.8cm}
\section{Frequency and Voltage Sensitivity Indicators}

In this section, the extraction of frequency and voltage sensitivity indicators, termed \ac{fsi} and \ac{vsi}, from the $PQ-V\omega$ transfer functions of \eqref{eq.pq_vw_matrix} is presented, which is the main contribution of this work.
The indices are computed in the frequency domain, revealing the active and reactive power to voltage and frequency relation across the network and across various 
%frequency ranges.
%
%and
dynamic phenomena timescales.

\subsubsection{Singular Value Decomposition}

For the derivation of these indices, the \ac{svd} is used~\cite{levine2018control}.
\ac{svd} decomposes a given complex-valued matrix into its singular values, introducing the following matrices:
\begin{equation}
\boldsymbol{U}\,\boldsymbol{\Sigma}\,\boldsymbol{V}^{\mathsf{H}}
=
\left[
\begin{smallmatrix}
| & & |\\
\boldsymbol{u}_1 & \cdots & \boldsymbol{u}_m\\
| & & |
\end{smallmatrix}
\right]
\left[
\begin{smallmatrix}
\sigma^{\max} & & \\
& \ddots & \\
& & \sigma^{\min}
\end{smallmatrix}
\right]
\left[
\begin{smallmatrix}
| & & |\\
\boldsymbol{v}_1 & \cdots & \boldsymbol{v}_n\\
| & & |
\end{smallmatrix}
\right]^{\mathsf H}.
\end{equation}
where $\boldsymbol{V}$ and $\boldsymbol{U}$ are the input and output complex orthonormal matrices, respectively, the superscript $(\cdot)^{\mathsf{H}}$ indicates the Hermitian matrix operator (conjugate transpose), and $\boldsymbol{\Sigma}$ is the singular value diagonal matrix.
The maximum and minimum singular values ($\sigma^{\max}$ and $\sigma^{\min}$) represent the maximum and minimum potential amplification, respectively.
Moreover, $\boldsymbol{v}_1$ and $\boldsymbol{u}_1$ are the worst input and output directions~\cite{dsi_onur}.
\vspace{-0.55cm}
\subsection{Frequency Sensitivity Indicator}

The proposed \ac{fsi} index is computed by taking the \ac{svd} of each row of the $\boldsymbol{G}_{\omega,P}(s)$ block matrix from~\eqref{eq.pq_vw_matrix}, as in:
\begin{equation}
\text{SVD}(
\underbrace{
    \begin{bmatrix}
        \boldsymbol{G}_{\omega_1,P_n}(s)
        &
        \dots
        &
        \boldsymbol{G}_{\omega_m,P_n}(s) \\
    \end{bmatrix}}_{\boldsymbol{G}_{\omega,P_n}(s)}),
\end{equation}
where $\boldsymbol{G}_{\omega,P_n}(s)$ is the transfer function from the active power injection at the $n$-th bus to the rotor speeds of all generation units. 
The complex matrix can be evaluated at a certain frequency point ($f$) from $\boldsymbol{G}_{\omega,P_n}(s)$ by setting $s=j\omega=j2\pi f$, resulting in:
\vspace{-0.1cm}
\begin{equation}
\boldsymbol{G}_{\omega,P_n}(f) = \boldsymbol{U}_{\omega,P_n} \, \boldsymbol{\Sigma}_{\omega,P_n} \, \boldsymbol{V}_{\omega,P_n}^{\mathsf{H}},
\end{equation}
where $\boldsymbol{U}_{\omega,P_n} \in \mathbb{C}^{m \times m}$ contains the left singular vectors associated with the output directions, $\boldsymbol{V}_{\omega,P_n} \in \mathbb{C}^{1 \times 1}$ is the right singular vector corresponding to the scalar input $\Delta P_n$, and $\boldsymbol{\Sigma}_{\omega,P_n}$ is the diagonal singular value matrix.

Since the input dimension is one, $\boldsymbol{\Sigma}_{\omega,P_n}$ reduces to a single singular value $\sigma^{\max}_{\omega,P_n}$, which quantifies the worst-case gain from an active power injection at Bus~$n$ to the generator frequency responses. The associated left singular vector $\boldsymbol{u}_{\omega,P_n}$ describes how this maximum gain is distributed among the generator frequencies, indicating the participation of each generator in the dominant response. Therefore, the proposed frequency sensitivity indicator is defined as
\vspace{-0.1cm}
\begin{equation} \label{eq.fsi_wp}
    \text{FSI}_{\omega,P_n}(f) = \left|\boldsymbol{u}_{\omega,P_n}\right| \sigma^{\max}_{\omega,P_n},
\end{equation}
where $\left|\boldsymbol{u}_{\omega,P_n}\right|$ denotes the element-wise absolute value of the dominant left singular vector. Repeating this analysis for every bus provides a network-wide measure of how strongly an active power disturbance at each location propagates through the system and excites the generator frequencies.

Equation~\eqref{eq.fsi_wp} combines the maximum amplification with the corresponding worst output direction, yielding the sensitivity of each generator and IBR frequency to an active power injection at Bus~$n$. Repeating this computation for every bus produces a network-wide frequency sensitivity map, highlighting the locations from which active power disturbances have the greatest influence on the system frequency dynamics.

The same procedure can be repeated considering reactive power injections. In that case, the \ac{fsi} can be written as:
\vspace{-0.1cm}
\begin{equation} \label{eq.fsi_wQ}
    \text{FSI}_{\omega,Q_n}(f) = \left|\boldsymbol{u}_{\omega,Q_n}\right| \sigma^{\max}_{\omega,Q_n},
\end{equation}
which quantifies the effect of reactive power variations on the rotor speeds of the generators.
%—a relationship that is non-negligible for resistive and coupled systems, such as low-inertia power systems with high \ac{ibr} penetration.
\vspace{-0.2cm}
\subsection{Voltage Sensitivity Indicator}

Following a similar procedure, the \ac{vsi} can be extracted by selecting the voltage magnitudes as outputs, leading to the \ac{vsi} for active power injections:
\vspace{-0.1cm}
\begin{equation} \label{eq.fsi_VP}
    \text{VSI}_{|V|,P_n}(f) = \left|\boldsymbol{u}_{|V|,P_n}\right| \sigma^{\max}_{|V|,P_n},
\end{equation}
and for reactive power injections:
\vspace{-0.1cm}
\begin{equation} \label{eq.fsi_VQ}
    \text{VSI}_{|V|,Q_n}(f) = \left|\boldsymbol{u}_{|V|,Q_n}\right| \sigma^{\max}_{|V|,Q_n}.
\end{equation}
\vspace{-0.85cm}
\subsection{Indicator Interpretation}

An important property of the proposed indices is their inequality interpretation. Values below 1 indicate that oscillatory active or reactive power injections are attenuated by the linear system, whereas values above 1 indicate output amplification and, therefore, \emph{high-sensitivity} conditions. Throughout this paper, the latter are used to identify operating conditions where disturbances are more strongly propagated through the system.

Since \ac{fsi} and \ac{vsi} are derived from the maximum potential amplification and the corresponding worst output direction, they are unbounded from above. Lower values indicate greater attenuation of oscillatory disturbances and, consequently, a more robust system response. However, values below 1 do not necessarily guarantee compliance with operational limits.

To account for operational security, acceptable deviations of $\pm1\%$ for frequency and $\pm10\%$ for voltage are adopted, consistent with standard power system security margins~\cite{YU2025101278}. Accordingly, $\text{FSI}_{\omega,P}$ or $\text{FSI}_{\omega,Q}$ values exceeding 0.01 indicate a critical frequency condition, whereas $\text{VSI}_{|V|,P}$ or $\text{VSI}_{|V|,Q}$ values exceeding 0.1 indicate a critical voltage condition.

\vspace{-0.3cm}

\subsection{Global Indices}
\label{sec.global_indices}

Beyond the frequency-based evaluation of the indices, a global characterization of the system dynamics can be established by defining the global indices $\overline{\text{FSI}}$ and $\overline{\text{VSI}}$ as:
\begin{equation}
\small
\label{eq.fsi_vsi_glo}
\overline{\text{FSI}}
=
\max_{f}\left\{\text{FSI}_{\omega,P},\,\text{FSI}_{\omega,Q}\right\},
\overline{\text{VSI}}
=
\max_{f}\left\{\text{VSI}_{|V|,P},\,\text{VSI}_{|V|,Q}\right\}.
\end{equation}
Equation~\eqref{eq.fsi_vsi_glo} provides an aggregated representation that enables the interpretation of the system-wide frequency and voltage behavior across the selected frequency ranges.
Moreover, if the global indices $\overline{\text{FSI}}$ and $\overline{\text{VSI}}$ are over 1, the system exhibits a highly sensitive frequency or voltage response. 
The global indices provide a first-level assessment by indicating whether the system exhibits high-sensitivity behavior. Subsequently, the frequency-based indices enable a deeper diagnosis by identifying the specific frequency ranges, buses, and power injections responsible for the observed behavior.
\vspace{-0.5cm}
\subsection{Relation Between FSI/VSI and Natural Modes}
\label{sec.fsi_vsi_natmodes}
\vspace{-0.1cm}
In this subsection, the mathematical relationship between the proposed indices and the structural natural modes of the system is established for both single- and multi-dominant pole conditions.
%
%Since the sensitivity indicators are derived directly from the \ac{svd} of the complex-valued $PQ\text{-}V\omega$ transfer matrix, elevated index values naturally manifest at frequencies where dominant oscillatory modes are excited. Consequently, the proposed indices do not merely reveal the existence of natural oscillations; they explicitly map how these dynamic disturbances propagate and spatially distribute throughout the transmission network.
%
The formal mathematical proof of this modal link is provided in Appendix~\ref{sec.mathlink}.
\subsubsection{Physical Meaning of $\sigma^{\max}$ via Left Eigenvectors ($\boldsymbol{\psi}_k$)}
The maximum singular value $\sigma^{\max}$ is proportional to the scalar product $\left|\boldsymbol{\psi}_k^{\top} \boldsymbol{B}_n\right|$, which isolates the $n$-th component of the $k$-th left eigenvector associated with the injection bus. Physically, the left eigenvector defines the modal controllability, quantifying how effectively an external disturbance at a specific input location can excite a given natural mode \cite{rogers_2000_power}. Therefore, a high local \ac{fsi}/\ac{vsi} value at Bus $n$ implies that the bus possesses high modal controllability. Disturbances injected at such nodes will strongly activate the corresponding poorly damped natural mode, making these buses highly sensitive vulnerability points in the grid.
\subsubsection{Physical Meaning of $\boldsymbol{u}$ via Right Eigenvectors ($\boldsymbol{\phi}_k$)}
The worst output direction vector $\boldsymbol{u}$ is proportional to the scalar product $\left|\boldsymbol{C}_m \boldsymbol{\phi}_k\right|$, which extracts the $m$-th component of the $k$-th right eigenvector observed at the output variables (e.g., generator rotor speeds). The right eigenvector defines the modal observability and structural mode shape, dictating how an excited natural mode distributes its energy across the system states and outputs \cite{rogers_2000_power}. Hence, a high index value at output $m$ indicates high modal participation of that specific generator or bus in the oscillation. At the resonance frequency, the spatial distribution of the sensitivity indicator converges to the physical mode shape of the network, enabling direct identification of the dominant oscillation pattern from standard frequency-domain measurements.

When a single natural mode ($\lambda_r$) is significantly closer to the imaginary axis or more highly excited than all other system poles at a given frequency of interest $f$, the multi-variable frequency response simplifies to a single-eigenvector approximation. Under this single-mode dominance condition, the specific impact of an active power injection at Bus $n$ on the generator rotor speed at Bus $m$ is explicitly quantified by the localized index:
\vspace{-0.1cm}
\begin{equation} \label{eq.fsi_single_mode_decoupled}
\text{FSI}_{m,n}(f)
=
\frac{
\left|\boldsymbol{C}_m \boldsymbol{\phi}_r\right|
\left|\boldsymbol{\psi}_r^{\top}\boldsymbol{B}_n\right|
}
{\left|j2\pi f-\lambda_r\right|},
\end{equation}
where $\boldsymbol{C}_m$ isolates the row corresponding to output $m$, and $\boldsymbol{B}_n$ isolates the column corresponding to input $n$. 

Equation \eqref{eq.fsi_single_mode_decoupled} shows that the sensitivity indicator is governed by a fundamental trade-off between spatial alignment (the numerator) and frequency proximity (the denominator). If no system poles are located near the excitation frequency $f$, the Euclidean distance $|j2\pi f - \lambda_r|$ in the complex plane becomes large for all natural modes in the network. As a result, the denominator dominates, and the total modal residue contribution is heavily attenuated, yielding a very low overall $\text{FSI}_{m,n}(f)$ value. Physically, this represents a ``stiff'' operating condition: the grid inherently attenuates power fluctuations at that specific frequency, preventing the propagation of localized disturbances into wider network-wide oscillations.

Conversely, in real power grids
%characterized by low rotational inertia and high penetrations of converter-interfaced generation,
the dynamic response is rarely governed by a single isolated mode. Instead, multiple control and electromechanical poles often cluster within similar narrow frequency bands. Under these multimodal conditions, individual modal truncation becomes invalid, and the indicator must account for the collective algebraic summation of all the states:
\begin{equation} \label{fsivsi_multimode}
\text{FSI}_{m,n}(f) = \left| \sum_{i=1}^{l} \frac{(\boldsymbol{C}_m \boldsymbol{\phi}_i) (\boldsymbol{\psi}_i^{\top} \boldsymbol{B}_{n})}{j2\pi f - \lambda_i} \right|,
\end{equation}
where $l$ is the total state dimension of the linearized system. 

The physical implication of the complex summation in \eqref{fsivsi_multimode} is profound, since the modal residues are complex-valued vectors containing both magnitude and phase information.
Consequently, in a multimodal system, the frequency at which the maximum operational amplification occurs does not necessarily coincide exactly with the imaginary part of any single natural mode ($\text{Im}\{\lambda_i\}$).
This phase interaction shifts the physical resonance peaks, especially if the mode is highly damped \cite{elia_damping}. 
%
%, illustrating why traditional eigenvalue analysis can misjudge the true worst-case vulnerability points and highlighting the necessity of the proposed frequency-domain \ac{fsi}/\ac{vsi} framework.
%Furthermore, when damping is high, the resonance frequency may slightly deviate from the natural mode frequency \cite{elia_damping}.

% Classical eigenvalue analysis describes the behavior of individual modes independently; however, it does not capture the aggregate amplification arising from modal interactions. In contrast, the proposed FSI/VSI formulation is derived from the singular value decomposition (SVD) of the transfer matrix and therefore inherently captures the combined effect of all contributing modes.
%
\vspace{-0.3cm}
\section{Application Methodology and Visualization}
This section first presents the two applications of the proposed indices and later describes the \ac{fsi} and \ac{vsi} visualization strategy that was selected for the case studies. 
%
%The presented topologies are used as representative examples of the two main converter control categories.
\vspace{-0.4cm}
\subsection{Two Applications of the Proposed Indices}
%\begin{itemize}
    %\item \textbf
    \subsubsection{Application 1: Natural oscillation detection}
    % The $PQ\text{-}V\omega$ screening method excites the system dynamics over the specified frequency range. As a result, high FSI or VSI values
    % As a result, when the damping ratio of a mode is high, the resonance frequency may deviate from the natural mode frequency, leading to elevated FSI or VSI values. 
    %The $PQ\text{-}V\omega$ screening method excites the system across the frequency range of interest and captures the resulting amplification behavior.
    By analyzing the frequency points associated with high FSI and VSI values, the aggregated effect of the dominant modes of the system can be identified. 
    %Therefore, the proposed indices can also provide
    %a frequency-domain indication of the system eigenvalues.
    The correspondence between high FSI/VSI regions and the natural modes of the system further validates the effectiveness of the proposed screening methodology. This mathematical relationship is explained in Section~\ref{sec.fsi_vsi_natmodes}.
    %\item \textbf{
    \subsubsection{Application 2: Analysis of oscillation propagation}
    
    By comparing the index values of different output variables for the same active or reactive power injection at a given bus, the propagation of disturbances through the network can be assessed. In particular, FSI or VSI values greater than 1 indicate disturbance amplification. If this occurs for multiple variables at the same frequency, the disturbance is propagating through the system.
    
%\end{itemize}

\vspace{-0.3cm}
\subsection{Index Visualization}

For the case studies, the proposed indices are visualized by means of 3-D heatmaps, whose structure is explained in this subsection.
In these heatmaps, the y-axis represents the $\Delta P$ or $\Delta Q$ injection at each bus, the x-axis corresponds to the network variable under study (rotor speed of the generation units or the voltage magnitude of each bus), and the z-axis indicates the oscillatory frequency of the disturbance.
Finally, the color scale denotes the magnitude of the corresponding index. 
In order to enhance readability, very low values of the indices are omitted while all values are plotted on a logarithmic scale.

The upper limit of each heatmap corresponds to the maximum value of each respective index ($FSI_{\omega,P_n}$, $FSI_{\omega,Q_n}$, $VSI_{|V|,P_n}$ or $VSI_{|V|,Q_n}$) for each case, indicating a distinct maximum value and range for each heatmap.
%that each heatmap may have a different upper range.
%
%
Since an index value
%$FSI_{\omega,P_n}$, $FSI_{\omega,Q_n}$, $VSI_{|V|,P_n}$ and $VSI_{|V|,Q_n}$ values
greater than 1 indicates disturbance amplification, a threshold of 0 (i.e., $\log_{10}(1)$) is used in the color bar.
In other words, in the 3-D plots, values below 0 represent secure conditions, while values above 0 signify a potential amplification of the studied disturbance.
%
%
%
%
%The frequency range of the oscillatory disturbances under study is
%selections are
%entirely user-dependent and can be selected according to the timescale of interest.
%
In this work, for the $FSI_{\omega,P}$ and $FSI_{\omega,Q}$, a frequency range of 0-50 Hz is selected, while for the $VSI_{|V|,P}$ and $VSI_{|V|,Q}$ a range of 0-300 Hz is selected,
%
%This selection is
motivated by the well-known timescale separation in power systems~\cite{van2007voltage}.

\vspace{-0.35cm}
\section{Validation Results}
\subsection{General Description of Case Studies}
In this work, two case studies are presented based on modified versions of the well-known IEEE 68-bus benchmark system.
In the first case study, the proposed indices are evaluated on the IEEE 68-bus system, comprising an SG-dominant generation mix with GFM and GFL converters. First, the impact of forced oscillations on the critical system variables is analyzed, demonstrating that the proposed indices accurately represent the amplification observed in the time-domain responses. Subsequently, the first application of the proposed indices, namely natural oscillation detection, is demonstrated by relating the dominant FSI and VSI peaks to the system natural modes.
%
%The objective is to demonstrate the impact of forced oscillations on generator frequencies and bus voltages using the proposed indices. Detection of forced oscillations is not addressed; instead, the focus is on quantifying their impact on system variables. Additionally, the first application of the proposed indices, i.e., the natural mode identification, is demonstrated by comparing the indices obtained via SVD and~\eqref{fsivsi_multimode}.
%
In the second case study, demonstrating the second application of the proposed indices, the benchmark network is further modified to a power system fully operated by \ac{ibr}-based generation, where GFM converters are responsible for providing voltage and frequency regulation.
Based on this fully \ac{ibr}-based example, two scenarios are considered. The first considers the full dynamics of the considered network, corresponding to an EMT modelling approach, while the second eliminates some of the dynamic effects found in the network, corresponding to an RMS system representation~\cite{cheah2026new}.
For all case studies, the generator modelling framework is similar to the one used in~\cite{dsi_onur,wp_Onur}.
%
%all generation units are modeled in the $qd$ reference frame and in per-unit. The GFM converters employ droop-based synchronization and include cascaded voltage and current controllers, along with a reactive power–voltage droop. The GFL converters operate in PQ control mode with droop-based grid support and synchronize through a PLL. Both converter models use an $LCL$ filter, an averaged (non-switching) representation, and assume a constant DC link. Further modeling details can be found in~\cite{dsi_onur}. Synchronous generators are modeled using the round-rotor GENROU model, as described in~\cite{68bus} equipped with an exciter model of AC4A and a governor model of IEEEG1.
%
The linear model derivation, as well as all the metric calculations presented in this work were performed using STAMP~\cite{arevalo2025matlab, STAMP_Github}.

% In both case studies, the analysis begins with observing global indices for the whole system and for the full frequency spectrum of interest, and continues with the frequency-based and per-variable indices, which are represented in 3-D FSI and VSI heatmaps. 
% %
% %Such heatmaps are used to identify the most critical frequency ranges and network locations where ill-conditioned system dynamics occur.
% %
% By analyzing the frequency-based indices, the propagation of an input oscillatory disturbance at a given bus can be identified.
% %
% Propagation is detected when, for a given power injection at a specific bus, multiple system variables exhibit index values greater than 1. Conversely, if all global indices remain below 1, the system is considered to not propagate inputs at any frequency.
% %
% Guided by these observations, targeted active and reactive power disturbances are applied to selected buses at the identified frequencies.
\vspace{-0.4cm}
\subsection{Case Study 1: SG–Dominant System with EMT model}
\begin{figure}[htbp]
    \centering
    \includegraphics[width=0.85\columnwidth]{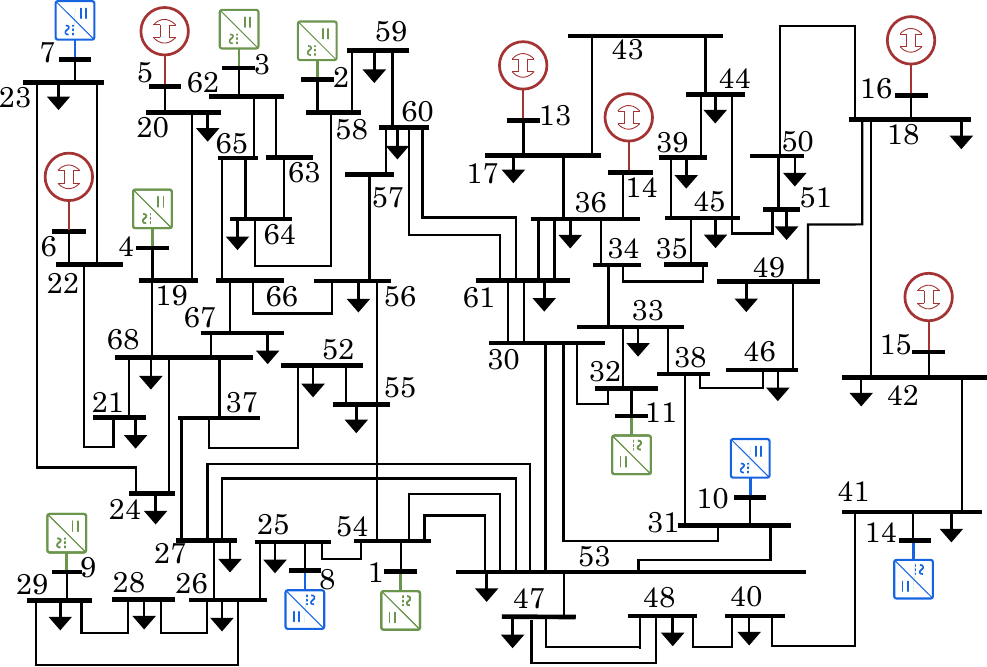}
    %\vspace{-0.1cm}
   % \captionsetup{font=footnotesize}
      \vspace{-0.25cm}
    \caption{Case Study 1: The modified IEEE 68 bus system. 4 GFL in blue, 6 GFM in green, 6 SG in red. }
    \label{fig:68_bus}
\end{figure}

% %
% In this section, an SG-dominant system is analyzed in the IEEE 68 bus system.
% %
% The objective of this case study is to demonstrate the effect of forced oscillations on generator frequencies and bus voltages using the proposed indices. 
% %
% The detection of forced oscillations is not considered; instead, the focus is on evaluating their impact on system variables.
% %
% Additionally, the second application of the proposed indices, natural mode identification, is demonstrated by comparing the indices obtained via SVD and Equation \eqref{fsivsi_multimode}. 
%
Fig.~\ref{fig:68_bus} shows the modified IEEE 68-bus system, which includes GFM (green) and GFL (blue) converters, as well as SG units (red).
Buses 1-4, 9, and 11 are equipped with GFM converters, while buses 7, 8, 10, and 14 host GFL converters.
Synchronous generators are installed at buses 5, 6, 13-16.
The overall generation mix consists of 65.2\% SG, 23.3\% GFL, and 11.5\% GFM units.

\begin{figure}[htbp]
    \centering
    \includegraphics[width=1\linewidth]{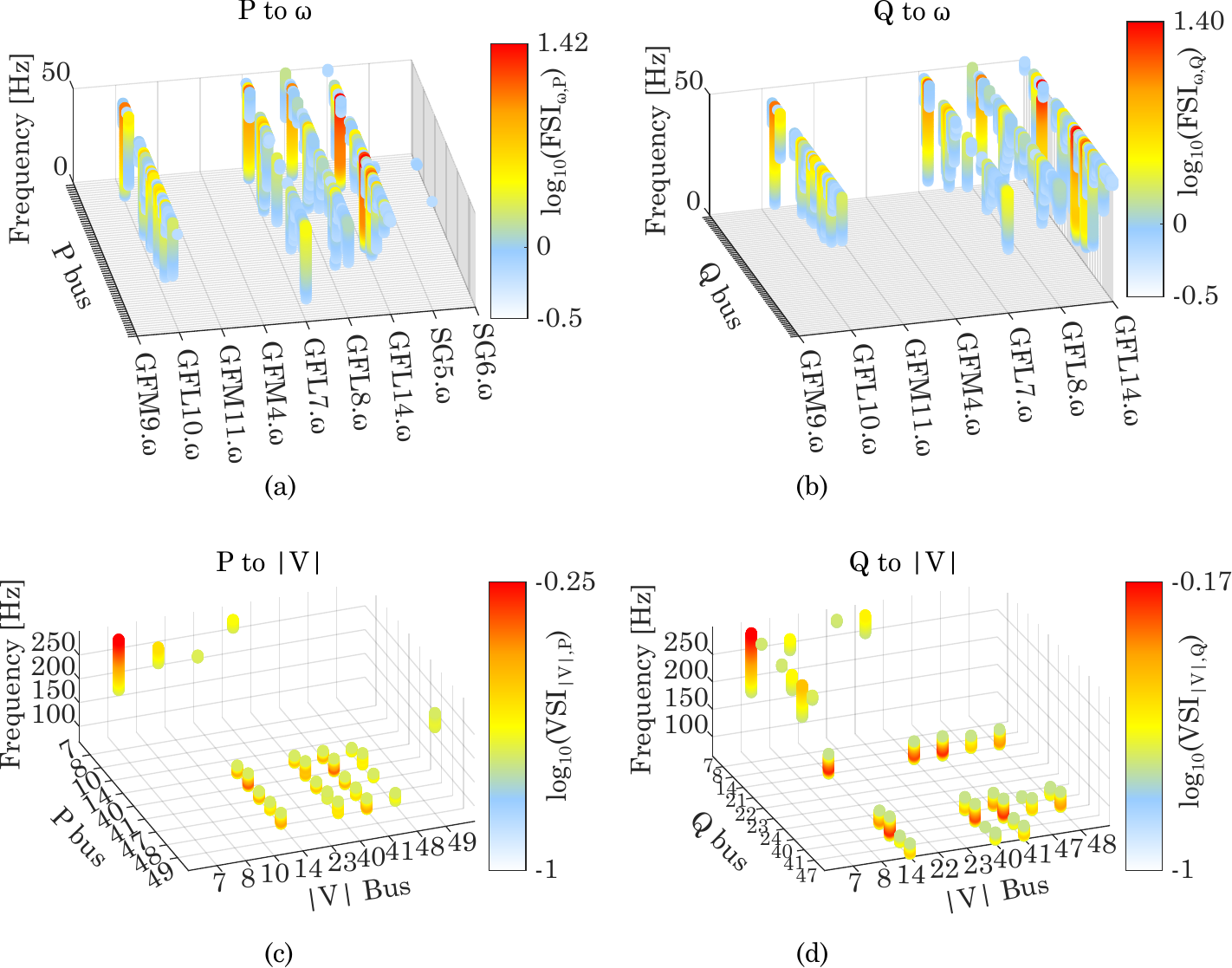}
   % \captionsetup{font=footnotesize}
      \vspace{-0.5cm}
\caption{3-D heatmaps of Case Study~1: (a) $P$--$\omega$ and $FSI_{\omega,P}$, (b) $Q$--$\omega$ and $FSI_{\omega,Q}$, (c) $P$--$|V|$ and $VSI_{|V|,P}$, and (d) $Q$--$|V|$ and $VSI_{|V|,Q}$.}
    \label{fig:68_scearnio1}
\end{figure}

By inspecting the global indices $\overline{\mathrm{FSI}}$ and $\overline{\mathrm{VSI}}$, it can be seen that the global indices present values of $\overline{\mathrm{FSI}}=26.3$ and $\overline{\mathrm{VSI}}=0.67$.
From the above values, it can be deduced that the frequency of the system is highly sensitive, according to the interpretation of  Section~\ref{sec.global_indices}, while the voltage remains within the critical operating margins.
Based on this initial screening of the whole system, further analysis is conducted on a bus-specific basis with the frequency-based indices.

Fig.~\ref{fig:68_scearnio1} shows the 3-D heatmaps of the  FSI and VSI metrics for the network of Case Study 1, with Figures~\ref{fig:68_scearnio1}(a) ands~\ref{fig:68_scearnio1}(b) show the $FSI_{\omega,P}$ and $FSI_{\omega,Q}$ values, respectively.
It can be seen that buses to which GFL units are connected exhibit significantly higher sensitivity to input disturbances than GFM and SG units, as reflected by their high $FSI_{\omega,P}$ and $FSI_{\omega,Q}$ values, while GFM and SG buses show very low sensitivity. In particular, GFL14 appears to be especially affected due to its electrical distance from the remaining units. 
%
%This behavior arises because GFL converters synchronize through a PLL and directly track the grid voltage; as a result, they follow incoming network disturbances, causing their internal frequency to be more strongly affected.
%
%In contrast, SG and GFM units possess the capability to impose their own frequency, enabling them to maintain a much stiffer frequency response under disturbances.
%
Moreover, the $FSI_{\omega,P}$ and $FSI_{\omega,Q}$ analyses provide complementary information: the active power-based index highlights the influence of active power interactions on frequency dynamics, while the reactive power-based index reveals the contribution of reactive power disturbances, allowing the dominant coupling mechanism at each bus to be identified.
% It can be observed that the frequencies of the GFL units are significantly more sensitive to input disturbances than those of the GFM and SG units, as indicated by the high $FSI_{\omega,P}$ and $FSI_{\omega,Q}$ values for GFL buses and the very low values for GFM and SG buses.
%
%This behavior arises because GFL converters synchronize through a PLL and directly track the grid voltage; as a result, they follow incoming network disturbances, causing their internal frequency to be more strongly affected.
%
%In contrast, SG and GFM units possess the capability to impose their own frequency, enabling them to maintain a much stiffer frequency response under disturbances.
%

Figures~\ref{fig:68_scearnio1}(c) and~\ref{fig:68_scearnio1}(d) present the $VSI_{|V|,P}$ and $VSI_{|V|,Q}$ indices.
All voltage-based VSI values remain below 1, indicating there are no highly sensitive voltage dynamics.
However, as the index values approach 0 in the 3-D heatmap plots, critical operation can arise.
For example, Fig.~\ref{fig:68_scearnio1}(c) reveals frequency-dependent vulnerabilities: Bus~7 is most sensitive to active power disturbances at 200--250~Hz, while buses~14, 40, 41, and~48 approach to critical range around 100~Hz. This is mainly due to GFL units at buses~7 and~14 and the electrical proximity of buses~40, 41, and~48 to GFL-controlled nodes.
A similar trend appears in Fig.~\ref{fig:68_scearnio1}(d), where the voltage at bus~7 is strongly affected by reactive power disturbances at buses~21--23 near 300~Hz, driven by the GFL unit at Bus~7 and its electrical coupling with these buses.
% In both cases, all voltage-based VSI values remain below~1, indicating there is no ill-conditioned voltage dynamics. Nevertheless, Fig.~\ref{fig:68_scearnio1}(c) reveals frequency-based voltage vulnerabilities. 
% In particular, Bus~7 exhibits the highest voltage sensitivity when an active power disturbance is applied at Bus~7 within the 200--250~Hz frequency range. 
% Furthermore, the voltage dynamics at buses~14, 40, 41, and~48 exceed the assumed safe operating limits when active power disturbances are applied at these buses around 100~Hz. 
% This behavior is primarily attributed to the presence of \ac{gfl} converters at buses~7 and~14, as well as the electrical proximity of buses~40, 41, and~48 to GFL-controlled units. Consequently, the voltage dynamics at these buses become more susceptible to disturbances within specific frequency ranges.
% In Fig.~\ref{fig:68_scearnio1}(d), behavior similar to that observed in Fig.~\ref{fig:68_scearnio1}(c) can be identified. 
% In addition, the voltage magnitude at Bus~7 is significantly affected by reactive power disturbances applied at buses~21, 22, and~23 in the vicinity of 300~Hz. 
% This behavior is attributed to the presence of a GFL-controlled unit at Bus~7 and the electrical interconnection among buses~21, 22, and~23, which enables the propagation of disturbance effects toward Bus~7.

From the above results, it can be concluded that GFL converters are significantly more sensitive to oscillatory disturbances than SGs and GFM converters. This is because GFL converters track the grid frequency, whereas SGs and GFM converters actively establish it. Similarly, buses hosting or electrically close to GFL converters exhibit greater voltage sensitivity over certain frequency ranges, while the voltage-source behavior of SGs and GFM converters provides stronger voltage regulation and reduces disturbance propagation~\cite{dsi_onur}.

\begin{table}[t!]
   \vspace{-0.1cm}
    \centering
   % \captionsetup{font=footnotesize}
    \caption{ Computation of $\boldsymbol{VSI}_{\boldsymbol{|V|},Q_{14}}$ at 76 Hz with SVD and link to natural modes}
       \vspace{-0.25cm}
    %\vspace{-0.2cm}
    \label{tab:fsivsi2natural}
    % Compact spacing and readable font
    \setlength{\tabcolsep}{3pt} % narrower columns
    \renewcommand{\arraystretch}{1.1} % slight vertical spacing
    \large% <-- makes text larger than what \resizebox compresses
    \resizebox{0.9\columnwidth}{!}{ % don't shrink too much
    \begin{tabular}{ccccccc}
        \toprule
        \multicolumn{5}{c}{$\boldsymbol{VSI}_{\boldsymbol{|V|},Q_{14}}$ values at 76 Hz computed by SVD} \\
        \midrule
        $VSI_{|V_{14}|,Q_{14}}$ & $VSI_{|V_{40}|,Q_{14}}$ & $VSI_{|V_{41}|,Q_{14}}$ & $VSI_{|V_{47}|,Q_{14}}$ & $VSI_{|V_{48}|,Q_{14}}$ \\ 
        \midrule
        0.590 & 0.541 & 0.582 & 0.331 & 0.428\\
        \midrule
        \multicolumn{5}{c}{$\boldsymbol{VSI}_{\boldsymbol{|V|},Q_{14}}$ values at 76 Hz computed by Eq.\eqref{fsivsi_multimode}} \\
        \midrule 
       0.590 & 0.541 & 0.582 & 0.331 & 0.428\\
        \bottomrule
    \end{tabular}}
\end{table}

Using the proposed approach, the natural oscillation detection of the system can also be studied. For example, a high value of the $VSI_{|V|,Q_{14}}$ is both detected in the heatmaps, derived from~\eqref{eq.fsi_VQ} around 76 Hz, as well as alternatively  computed by using~\eqref{fsivsi_multimode}.
%
%This analysis servers to link the SVD with the natural modes of the system. 
%
Table~\ref{tab:fsivsi2natural} shows the $\boldsymbol{VSI}_{\boldsymbol{|V|},Q_{14}}$ computed at 76 Hz both through the proposed SVD-based method of~\eqref{eq.fsi_VQ}, as well as~\eqref{fsivsi_multimode}, showcasing identical values.
%It can be seen that, the computation of the index by the two different methods, i.e., with  provides the same value,
%
% The above validates the analysis of Section~\ref{sec.fsi_vsi_natmodes}. Since~\eqref{fsivsi_multimode} uses eigen-properties (right and left eigenvectors) and natural modes of the system, the $\boldsymbol{VSI}_{\boldsymbol{|V|},Q_{14}}$ computed through SVD can contain significant information about the natural modes of the system.

Modal analysis of the linearized system reveals a critical, poorly damped pole pair at $-38.2 \pm j468.2$, corresponding to an oscillation frequency of $74.5\text{ Hz}$ and a low damping ratio of $8\%$. Analysis of the structural participation factors indicates that the states from Buses 14, 40, 41, 47, and 48, along with the GFL unit at Bus 14, dominate this mode. 

These results directly validate the consistency of the proposed method. Specifically, the modal resonance at $74.5\text{ Hz}$ aligns precisely with the frequency range where the proposed \ac{fsi} predicts prominent, high-sensitivity cases. Furthermore, the spatial distribution of the worst output direction vector at this frequency reaches its maximum magnitudes at these exact same nodes, confirming that the proposed indices accurately capture both the frequency and the network propagation of the underlying natural oscillation.
\vspace{-0.3cm}
\subsection{Case Study 2: Fully Inverter-Based Resource System}
\begin{figure}[t!]
    \centering
    \includegraphics[width=0.85\linewidth]{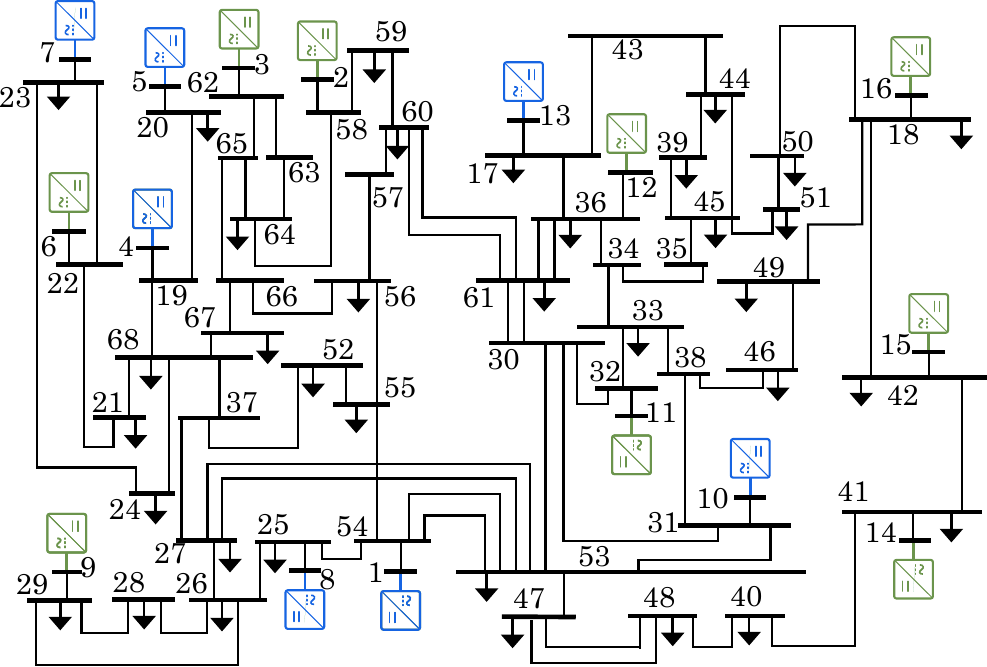}
   % \captionsetup{font=footnotesize}
      \vspace{-0.2cm}
    \caption{Case Study 2: IEEE modified 68 bus system with 9 GFM (in green) 7 GFL (in blue) converters}
    \label{fig:68network}
\end{figure}

Fig.~\ref{fig:68network} presents the modified IEEE 68 bus system for Case Study 2, with generation exclusively comprising GFM (in green) and GFL (in blue) converters.
The buses that are 1, 4, 5, 7, 8, 10, 13 are selected as GFL converter with a total GFL penetration in the generation mix of 30.3\%.

% In this section, a modified version of the IEEE 68-bus system, including GFM and GFL converters, is analyzed with the use of the proposed indicators in order to analyze the propagation of the signal which is the first application of the indeces.
% %The objective is to investigate how oscillations propagate through the network using the proposed FSI and VSI indicators. Two scenarios are considered.
% %
% %In order to investigate the oscillation propagation through the network using the proposed FSI and VSI indicators. Two scenarios are considered.
% %

% For the case study, two scenarios are considered.
% % The first represents a present-day power system dominated by SGs with some \ac{ibr} penetration.
% The first represents a future system with full \ac{ibr} penetration, where \ac{gfm} converters supply most of the generation and considering the network equations as differential equations, also known as EMT. The second scenario repeats the first scenario but by changing the network equations, to algebraic as known as RMS. 
% Finally, the third scenario repeats the second scenario using the RMS network model.
%
%
%

%
%In this case study, the proposed indices are validated by analyzing the propagation of active and reactive power oscillatory injections through time-domain simulations. The study demonstrates that the frequency-domain indicators accurately capture oscillation propagation while highlighting the influence of EMT and RMS modeling approaches.
%

\subsubsection{Scenario 1:IBR –Dominant System with EMT modeling}
\label{sec.case2sc1}
%
% \textcolor{red}{This section is a bit too long and chaotic. Try to introduce subsubsections, or more structure in general, and follow the order analysis, time domain plots with linear models, validation with nonlinear models.}
%
In this scenario, a 
%future
power system with a predominantly GFM-based \ac{ibr} generation mix is analyzed using the proposed FSI and VSI 3-D heatmaps. The frequency-based indices are used to assess and visualize oscillation propagation through geographical heatmaps, while EMT nonlinear simulations validate the obtained results.

\begin{figure}[t!]
    \centering
    \includegraphics[width=1\linewidth]{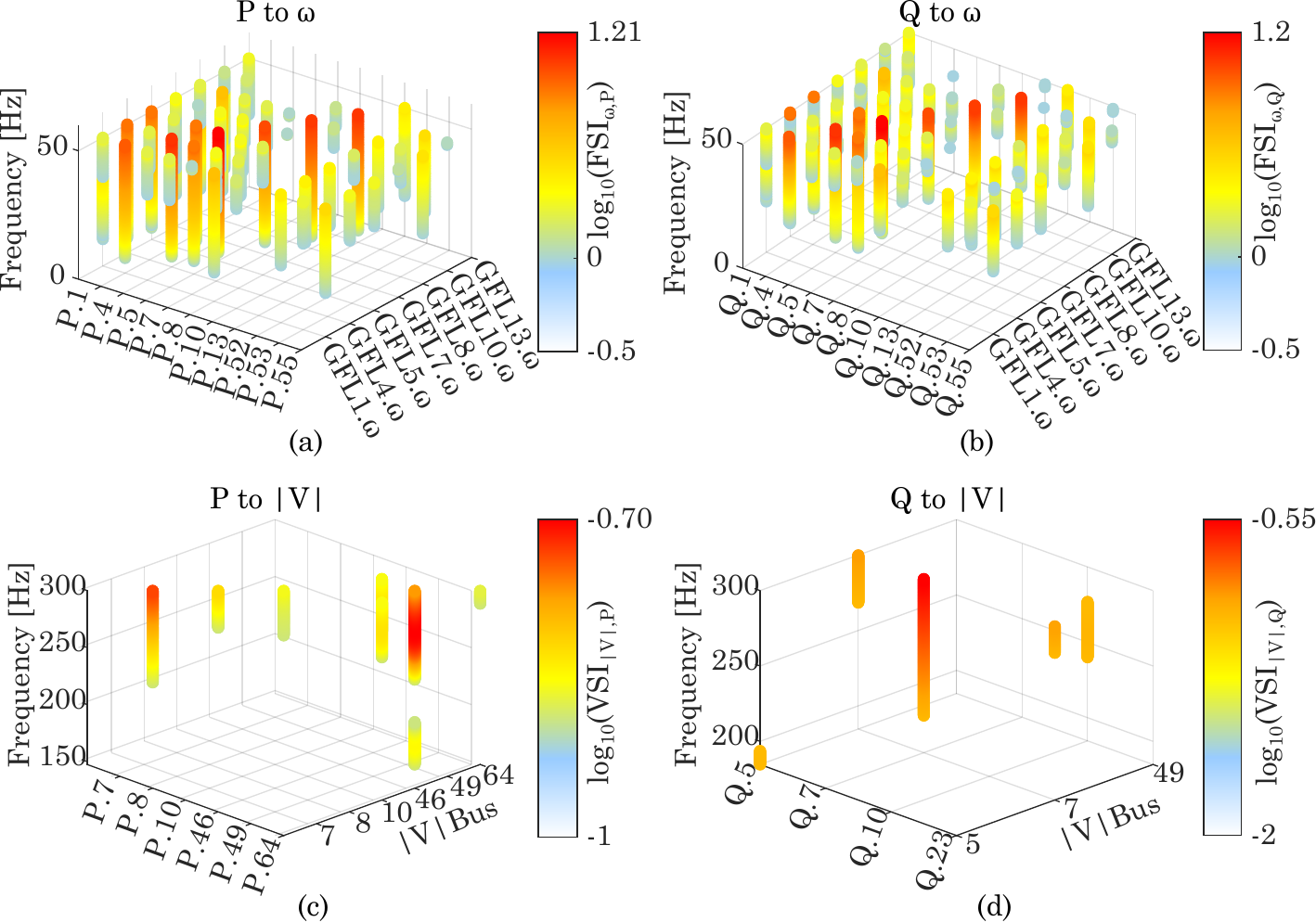}
   % \captionsetup{font=footnotesize}
      \vspace{-0.4cm}
\caption{3-D heatmaps of Case Study~2 Scenario 1: (a) $P$--$\omega$ and $FSI_{\omega,P}$, (b) $Q$--$\omega$ and $FSI_{\omega,Q}$, (c) $P$--$|V|$ and $VSI_{|V|,P}$, and (d) $Q$--$|V|$ and $VSI_{|V|,Q}$.}
    \label{fig:case2_heatmap}
\end{figure}

% In Fig.~\ref{fig:case2_heatmap}, the 3-D heatmaps of the PQ--V$\omega$ transfer functions obtained using the FSI and VSI are presented for Scenario~1.
%

By considering the global indices, $\overline{\mathrm{FSI}}$ and $\overline{\mathrm{VSI}}$, which showcase values 16.21 and 0.28, respectively, it is concluded that the system showcases highly sensitive frequency dynamics.
To gain further insight, frequency-dependent indices are analyzed.
Figures~\ref{fig:case2_heatmap}(a) and (b), showing $FSI_{\omega,P}$ and $FSI_{\omega,Q}$, indicate that GFL unit frequencies exhibit FSI values greater than 1 at buses 1, 4, 5, 7, 8, 10, 13, 52, 53, and 55 within the 40--50~Hz range.
$PQ$ injections are also observed to influence multiple GFL units, highlighting disturbance propagation across the network.

%
%
%Note that, other disturbance cases can also be selected. 
%
% Thus, these cases demonstrate an additional application of the proposed indicators: diagnosing how injected oscillations at specific buses and frequencies propagate through the system and influence both the generator rotor speeds and the voltage magnitudes at each bus.
%

Figures~\ref{fig:case2_heatmap}(c) and~\ref{fig:case2_heatmap}(d), presenting $VSI_{|V|,P}$ and $VSI_{|V|,Q}$, show that all VSI values remain below 1, indicating no amplification of the voltage response hence no propagation.
However, Fig.~\ref{fig:case2_heatmap}(c) reveals that voltage magnitudes at buses 7, 8, 10, 46, 49, and 64 exceed safe operating limits (10\%) under active power disturbances in the 150--300~Hz range, mainly due to GFL converters at buses 7, 8, and 10 and the electrical remoteness of buses 46, 49, and 64.
Similarly, Fig.~\ref{fig:case2_heatmap}(d) shows that buses 5, 7, and 49 violate safe limits under reactive power disturbances, attributed to GFL converters at buses 5 and 7 and the electrical distance of bus 49 from main generation sources.

In order to further illustrate the significance of the performed analysis to the dynamic operation of the system and to the system response in the time domain, two disturbance cases are selected for further analysis.
Due to their significant impact, as identified from the 3-D FSI and VSI heatmaps, on electrically distant GFL units.: a $P_{55}$ injection at 42~Hz and a $Q_{8}$ injection at 47~Hz are selected.
%
%First checking global and later frequency-based indices, the propagation of the signal is detected in certain disturbances. So, analysis will be navigated through that disturbances and their effect on propatation.
%

\begin{table}[t!]
   \vspace{-0.1cm}
    \centering
   % \captionsetup{font=footnotesize}
    \caption{FSI Values for Selected Cases }
       \vspace{-0.25cm}
    %\vspace{-0.2cm}
    \label{tab:fsivalues}
    % Compact spacing and readable font
    \setlength{\tabcolsep}{3pt} % narrower columns
    \renewcommand{\arraystretch}{1.1} % slight vertical spacing
    \large% <-- makes text larger than what \resizebox compresses
    \resizebox{0.9\columnwidth}{!}{ % don't shrink too much
    \begin{tabular}{ccccccc}
        \toprule
        \multicolumn{7}{c}{$FSI_{\omega,P_{55}}$ values at 42 Hz} \\
        \midrule
        $FSI_{\omega_1,P_{55}}$ & $FSI_{\omega_4,P_{55}}$ & $FSI_{\omega_5,P_{55}}$ & $FSI_{\omega_7,P_{55}}$ & $FSI_{\omega_8,P_{55}}$ & $FSI_{\omega_{10},P_{55}}$ & $FSI_{\omega_{13},P_{55}}$ \\
        \midrule
        2.52 & 1.42 & 1.4 & 0.86 & 2.24 & 0.75 & 0.22 \\
        \midrule
        \multicolumn{7}{c}{$FSI_{\omega,Q_8}$ values at 47 Hz} \\
        \midrule
        $FSI_{\omega_1,Q_8}$ & $FSI_{\omega_4,Q_8}$ & $FSI_{\omega_5,Q_8}$ & $FSI_{\omega_7,Q_8}$ & $FSI_{\omega_8,Q_8}$ & $FSI_{\omega_{10},Q_8}$ & $FSI_{\omega_{13},Q_8}$ \\
        \midrule
        4.2& 1.93 & 1.9 & 1.27 & 10.4 & 1.33 & 0.46 \\
        \bottomrule
    \end{tabular}}
\end{table}

Table~\ref{tab:fsivalues} shows the $FSI_{\omega,P_{55}}$ and $FSI_{\omega,Q_8}$ values without the logarithmic scale for each GFL unit at 42~Hz and 47~Hz, respectively.
For $FSI_{\omega,P_{55}}$, the GFL units located at buses 1, 4, 5, and 8 exhibit values greater than 1, indicating that an active power injection at Bus~55 propagates to these units and amplifies their response to the respective oscillations.
Similarly, the GFL units at buses 1, 4, 5, 7, 8, and 10 present $FSI_{\omega,Q_8}$ values greater than 1, reflecting that a reactive power injection at Bus~8 leads to frequency amplification in these units.

%
% The reason behind that the selection of such bus and specific frequencies is to illustrate the propagation and the effect of the applied disturbance.

\begin{figure}[t!]
    \centering
    \includegraphics[width=0.95\linewidth]{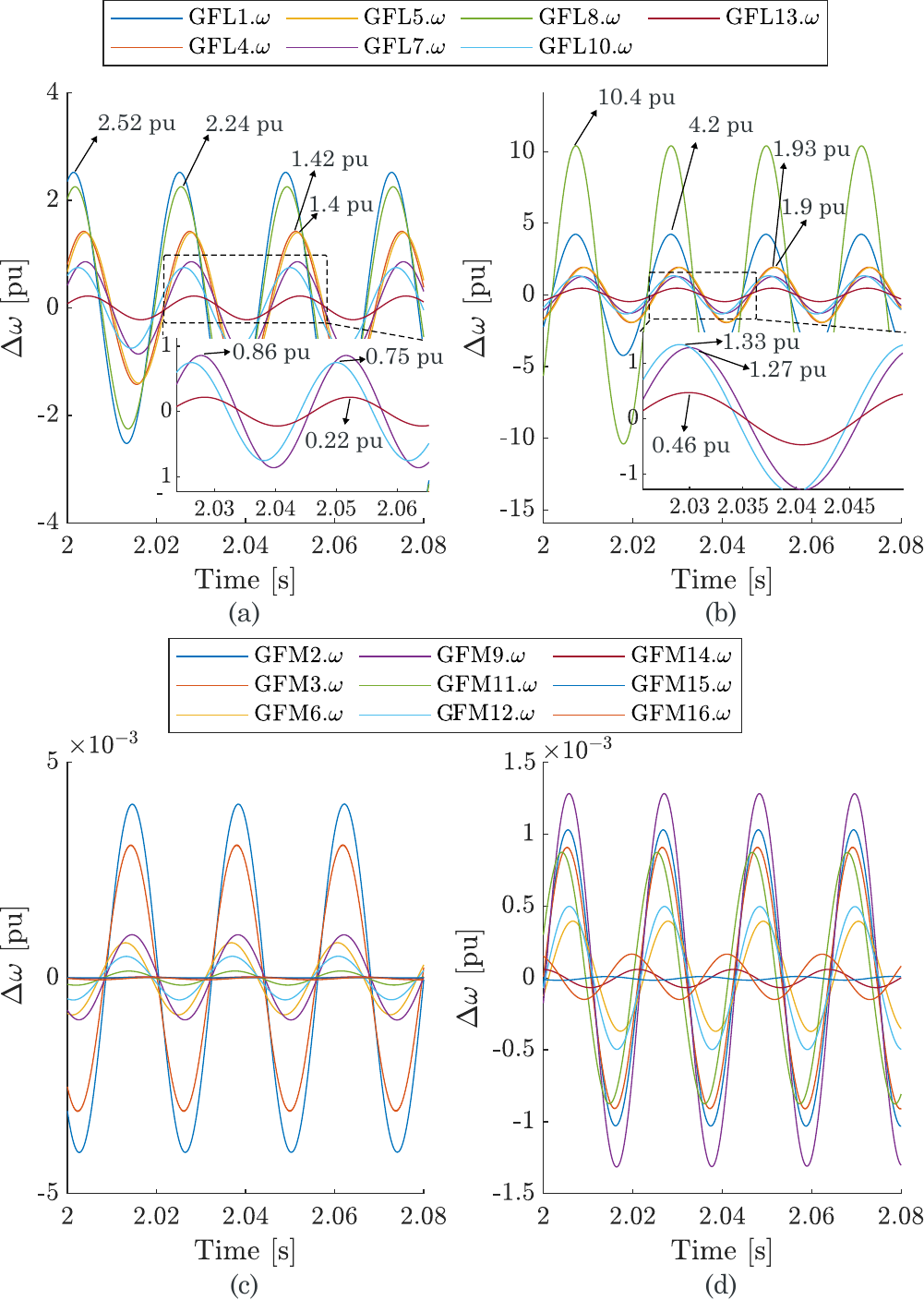}
   % \captionsetup{font=footnotesize}
      \vspace{-0.2cm}
\caption{Time-domain responses of GFL and GFM units in the linear model to unit sinusoidal disturbances: (a)–(b) GFL responses to P at 42 Hz (bus 55) and Q at 47 Hz (bus 8) respectively; (c)–(d) corresponding GFM responses}
    \label{fig:casestudy2_td}
\end{figure}
Fig.~\ref{fig:casestudy2_td}(a) shows the time-domain responses of power system linear model
%
%the frequency responses
%
of GFL units when a 1~pu sinusoidal $P$ disturbance of 42~Hz at Bus~55 applied.
It is noted that this high disturbance value is selected to highlight the variety of disturbance amplification values across the network, based on its topology and dynamic characteristics, which the proposed indices accurately capture.
Amplifications of 2.52, 1.42, 1.40, and 2.24~pu are observed at buses 1, 4, 5, and 8, respectively, while buses 7, 10, and 13 exhibit non-amplified responses (0.86, 0.75, and 0.22~pu).
Fig.~\ref{fig:casestudy2_td}(b) presents the response to a 1~pu, 47~Hz sinusoidal $Q$ disturbance at Bus~8, where buses 1, 4, 5, 7, 8, and 10 exhibit amplifications of 4.2, 1.93, 1.90, 1.27, 10.4, and 1.33~pu, respectively, while Bus~13 remains weakly affected (0.46~pu). Figs.~\ref{fig:casestudy2_td}(c)–(d) show that GFM frequency deviations remain below 0.01~pu under both disturbances, indicating safe operation. The measured time-domain amplification coincides with the proposed frequency-domain indices, validating the consistency of the methodology. Since the oscillatory reactive power disturbance is injected at Bus~8, the highest amplification (10.4 pu) occurs at the local GFL unit and progressively propagates to the remaining GFL units. Consequently, GFL~1, being electrically closest to the disturbance location, exhibits the second-largest response (4.2 pu), whereas the more electrically distant GFL units experience progressively lower amplification.
%
% , confirming that the proposed FSI and VSI capture disturbance propagation and its impact on generator frequency behavior.

\begin{figure}[t!]
    \centering
    \includegraphics[width=0.85\linewidth]{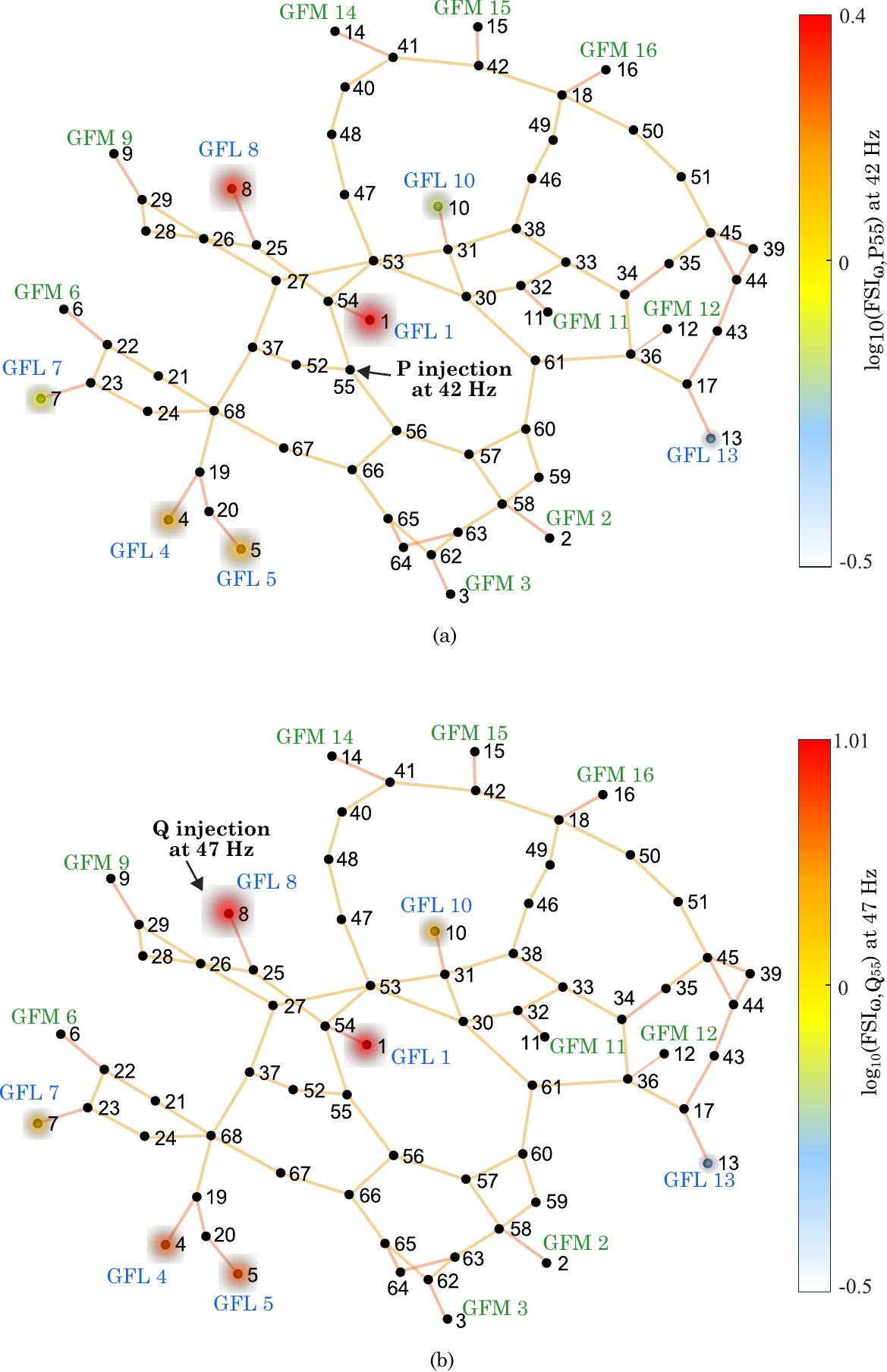}
   % \captionsetup{font=footnotesize}
      \vspace{-0.2cm}
    \caption{ Geographical heatmaps of IEEE modified 68 bus system (a) $FSI_{\omega,P_{55}}$ at 42 Hz (b) $FSI_{\omega,Q_{8}}$ at 47 Hz with EMT network}
    \label{fig:casestudy2_heatmap}
\end{figure}

The spatial characteristics of the proposed oscillation analysis are further illustrated through network heatmaps.
Fig.~\ref{fig:casestudy2_heatmap}(a) shows the response to a 42~Hz active power injection at Bus~55, where the disturbance mainly propagates to the nearby GFL units at buses 1 and 8, followed by buses 4 and 5, while buses 7, 10, and 13 exhibit progressively weaker responses.
Similarly, Fig.~\ref{fig:casestudy2_heatmap}(b) presents the response to a 47~Hz reactive power injection at Bus~8. The largest response occurs at buses 8 and 1, with noticeable propagation to buses 4, 5, 7, and 10, whereas Bus~13 remains largely unaffected due to its electrical remoteness.
In both cases, the GFM unit frequencies are largely insensitive to the disturbance location, as they establish rather than track the grid frequency, unlike GFL converters.
\begin{figure}[htbp]
    \centering
    \includegraphics[width=0.85\linewidth]{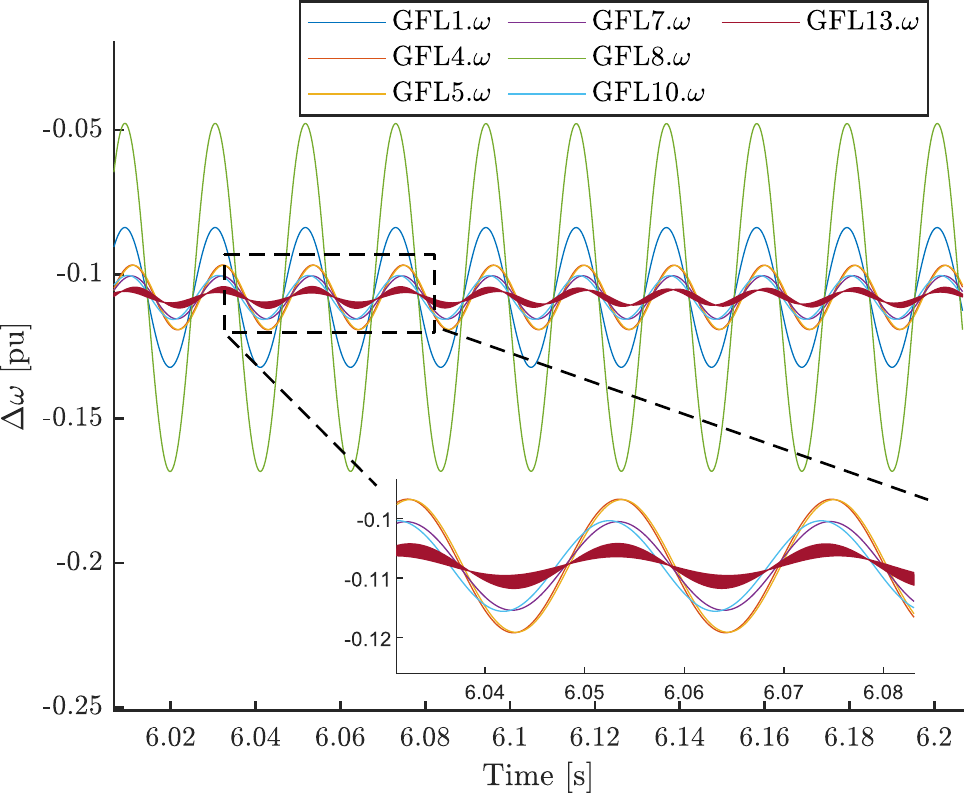}
   % \captionsetup{font=footnotesize}
      \vspace{-0.2cm}
    \caption{Time domain responses of the generation unit frequencies of Case Study 2, Scenario 1 Injection Q to bus 8 at 47 Hz in the nonlinear model}
    \label{fig:casestudy2_td_nonlinear}
\end{figure}

% %
% \textcolor{red}{Similar to the linear model, is there a validation only for scenario 2? Is there a reason for that?
% %
% Maybe we can just frame it in a different way altogether if you want to avoid additional figures.}
% %
%
Fig.~\ref{fig:casestudy2_td_nonlinear} represents the nonlinear model validation
%which is used to make sure that it is well understood, 
of Case Study 2, Scenario 1 for an indicative disturbance of reactive power, injected to bus 8 at 47 Hz.
For the validation with the nonlinear system, a similar test was performed with the one of the linear case, presented in Section~\ref{sec.case2sc1}.
%
%The same validation method that is explained in the previous section is applied.
%
The nonlinear time-domain response shows that GFL 8 experiences the largest frequency deviation, followed by GFL 1, GFL 4, GFL 5, GFL 10, GFL 7, and GFL 13.
The nonlinear simulation results show the same trend as observed in FSI plots and linear time-domain simulations, confirming the consistency of the proposed analysis.

The results demonstrate that the proposed frequency-based indices accurately diagnose oscillation propagation. Consistent with the time-domain responses, GFL converters are highly sensitive to injected disturbances and induce frequency vulnerabilities at electrically nearby buses, whereas GFM converters effectively mitigate these effects.
%
%This highlights the superior dynamic robustness provided by GFM operation in \ac{ibr}-dominated systems.
%

\subsubsection{Scenario 2: IBR–Dominant System with RMS modeling}
%
%
% \begin{figure}[t!]
%     \centering
%     \includegraphics[width=1\linewidth]{casestudy2/68RMS_manuscript.pdf}
%    % \captionsetup{font=footnotesize}
%       \vspace{-0.2cm}
% \caption{3-D heatmaps of Case Study~2 Scenario 2: (a) $P$--$\omega$ and $FSI_{\omega,P}$, (b) $Q$--$\omega$ and $FSI_{\omega,Q}$, (c) $P$--$|V|$ and $VSI_{|V|,P}$, and (d) $Q$--$|V|$ and $VSI_{|V|,Q}$.}
%     \label{fig:68_rms}
% \end{figure}
% In this scenario, the effect of neglecting the network dynamics on the oscillation propagation will be studied.
% %the scenario 1 will be studied without considering the network dynamics. 
% %
% %The objective is to observe the effect of converting line equations from differential to algebraic on the propagation of the oscillations. 
% %
% Fig.~\ref{fig:68_rms} presents the frequency-based indices are represented, with the maximum values of each index being reduced compared with the EMT case. 
% In Scenario 2, $\overline{\mathrm{FSI}}$ and $\overline{\mathrm{VSI}}$ have smaller values (9.33 and 0.033, respectively) compared to the EMT case, indicating that the RMS case might overlook propagating oscillations in the system.
%

In order to compare the RMS case with the EMT case, $FSI_{\boldsymbol{\omega},P_{55}}$ at 42 Hz is selected (same as Scenario 1), and the same upper and lower margins are kept for the comparison. 
\begin{figure}[!h]
    \centering
    \includegraphics[width=0.9\linewidth]{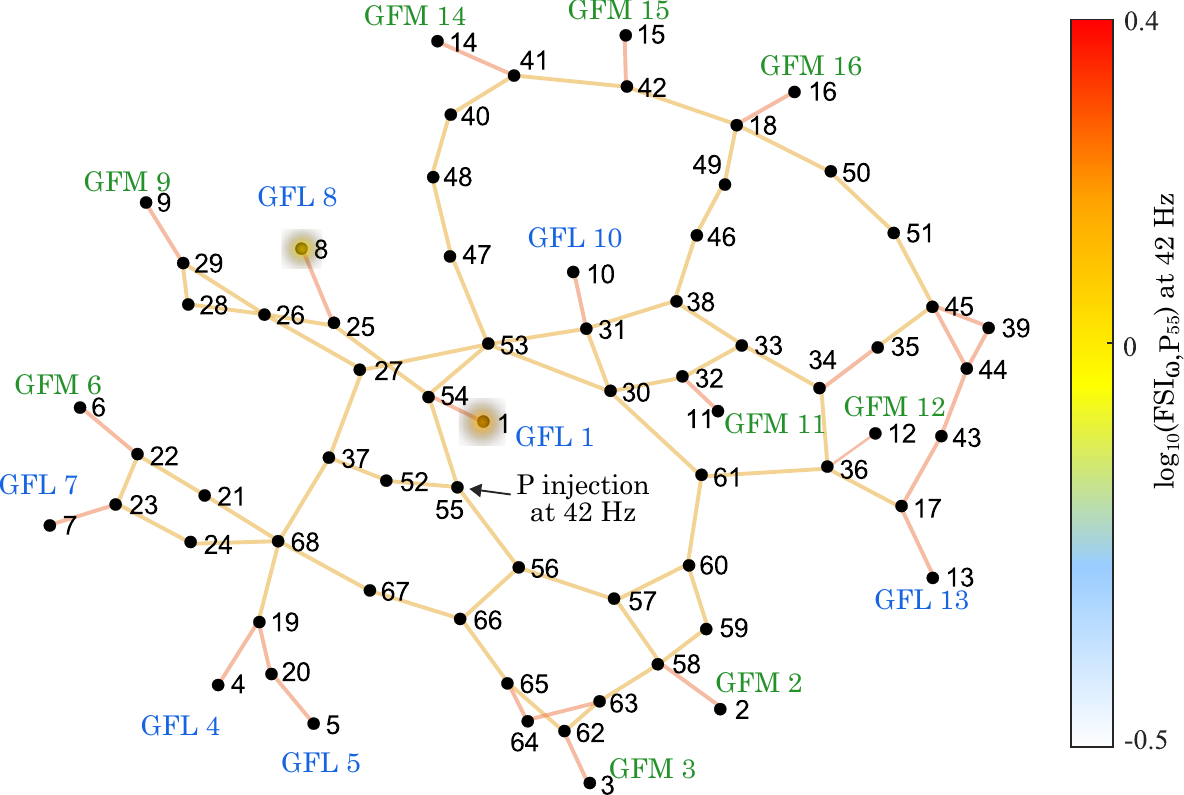}
   % \captionsetup{font=footnotesize}
   \vspace{-0.2cm}
    \caption{Geographical heatmap of Case Study 2, Scenario 2 with RMS network P injection at 42 Hz to bus 55.}
    \label{fig:68_rms_heatmap}
\end{figure}
Fig.~\ref{fig:68_rms_heatmap} illustrates the effect of $P$ injection at bus 55 with a frequency of 42 Hz on the frequency of each generation unit.
It can be observed that only GFL1 and GFL8 are affected, and the affect is significantly smaller compared to the EMT case.
The GFL units located at buses 4, 5, and 7 show no observable impact.
In contrast, EMT simulations provide a more accurate representation of oscillation propagation. Although RMS simulations are computationally more efficient for large-scale systems, this efficiency is achieved by neglecting network electromagnetic dynamics. As a result, RMS models may produce optimistic assessments and underestimate the propagation of oscillatory disturbances that could lead to critical operating conditions. The proposed FSI and VSI consistently capture this difference by indicating higher oscillation amplification for the EMT model.
\section{Conclusion}
\vspace{-0.1cm}
This work 
%presented a frequency-domain screening method with Power to Voltage and Frequency dynamics ($PQ–V\omega$ transfer functions) and 
introduces two quantitative indicators, namely the \ac{fsi} and the \ac{vsi},
that relate power to voltage and frequency dynamics for oscillation detection and propagation assessment.
%
% These metrics are mathematically linked to the eigenproperties of the system, so they can detect natural oscillations and assess oscillation propagation in networks with both \ac{sg} and \ac{ibr}-based penetration.
%
By analytically extracting the linear $PQ–V\omega$ transfer functions from EMT models, the proposed indices provide clear, localized insight into how power disturbances affect bus voltages and generator rotor speeds across multiple timescales.
Compared to previous approaches, the proposed framework does not consider voltage and frequency oscillations separately, but includes both using a unified approach. 
The distinction between global and frequency-based indices enables a two-step assessment: global indices allow rapid detection of variables susceptible to oscillations, while frequency-based indices support detailed diagnosis by identifying critical buses, frequency ranges, and disturbance locations.
Results from case studies based on the IEEE 68-bus benchmark show that FSI and VSI offer a practical and scalable screening method for operators to enhance system-level oscillation tracking and security assessment in increasingly inverter-dominated grids.
%
% As future work, control-level mitigation strategies based on the FSI and VSI will be investigated.
% %
% In addition, a methodology for selecting an appropriate reference model for grid integration of \acp{ibr} using the FSI and VSI will be developed.
 %hence the effect of injecting active power to bus 1 at any frequency can be observed with FSI.

% In \eqref{eq.pq_vw_matrix}, the block $G_{\omega,P_1}(s)$ represents the transfer function from the active power injection at Bus~1 to the frequency deviations of all generating units. Its singular value decomposition is written as

% or

\vspace{-0.2cm}
\appendices
\vspace{-0.2cm}
\section{Generation unit data}
\vspace{-0.2cm}
The controller parameters of the GFL and GFM converters are summarized in Table~\ref{table:controller_parameters}. The GFM outer-loop control parameters for Case Study~2 Scenarios~1 and~2 are given in Table~\ref{table:gfm_outer_param}, while the hardware parameters of both GFL and GFM converters are listed in Table~\ref{table.hardware_param}.
\vspace{-0.2cm}
\begin{table}[htbp]
\caption{GFM and GFL Hardware Parameters}
\vspace{-0.25cm}
\label{table.hardware_param}
%\scriptsize	
\centering
\begin{tabular}{|c|c|c|}
\hline
Parameter & Symbol & Value\\
\hline\hline
\makecell{Transformer Impedance} & $R_{tr}+jX_{tr}$ & 0.005+0.1~pu \\
\hline
\makecell{Converter Filter Impedance} & $R_{f}+jX_{f}$ & 0.005+0.1~pu \\
\hline
\makecell{Converter Filter Susceptance (GFM)} & $B_{f}$ & 0.15~pu \\
\hline
\makecell{Converter Filter Susceptance (GFL)} & $B_{f}$ & 0.01~pu \\
\hline
\end{tabular}
%\vspace{-0.3cm}
\end{table}
\begin{table}[htbp]

\caption{Controller Parameters for All Case Studies}
\vspace{-0.25cm}
\label{table:controller_parameters}
\centering
\begin{tabular}{|c|c|c|c|}
\hline
\textbf{Type} & \textbf{Parameter} & \textbf{Symbol} & \textbf{Value} \\
\hline\hline

\multirow{6}{*}{GFL}
& \makecell{Current and $PQ$\\ controller time constant}
& $\tau_{cc}$,$\tau_{PQ}$
& 1 ms, 1 s \\
\cline{2-4}

& \makecell{$f$--$P$ droop gain\\ and time constant}
& $m_{f-P}$, $\tau_{f-P}$
& 0.5, 0.1 s \\
\cline{2-4}

& \makecell{PLL time constant\\ and damping ratio}
& $\tau_{PLL}$, $\xi_{PLL}$
& 0.1 s, 0.707 \\
\cline{2-4}
\hline
\multirow{2}{*}{GFM}
& \makecell{Current control\\time constant}
& $\tau_{cc}$
& 1 ms \\
\cline{2-4}

& \makecell{AC voltage control\\time constant, damping ratio}
& $\tau_{vc},\,\xi_{vc}$
& 50 ms, 0.707 \\
\hline

\end{tabular}
\end{table}
\vspace{-0.75cm}
\begin{table}[htbp]
\caption{GFM Outer Loop Control Parameters}
\vspace{-0.25cm}
\label{table:gfm_outer_param}
\centering
\begin{tabular}{|c|c||c|c|}
\hline
\multicolumn{2}{|c||}{\textbf{Scenario 1}} &
\multicolumn{2}{c|}{\textbf{Scenario 2}} \\
\hline
\textbf{GFM Unit} &
\textbf{Time Constant} &
\textbf{GFM Unit} &
\textbf{Time Constant} \\
\hline\hline
GFM 7  & 0.22 & GFM 2            & 0.14 \\
GFM 8  & 0.25 & GFM 3, 15        & 0.19 \\
GFM 10 & 0.21 & GFM 6            & 0.60 \\
GFM 14 & 0.20 & GFM 9            & 0.25 \\
       &      & GFM 11, 14, 16  & 0.20 \\
       &      & GFM 12           & 0.22 \\
\hline
\end{tabular}
\end{table}
%

\section{Mathematical Link Between Natural Modes and FSI/VSI}

\label{sec.mathlink}

This appendix derives the relationship between the proposed sensitivity indicators and the state-space natural modes of the system.
For the linear time-invariant system
$\dot{\boldsymbol{x}}=\boldsymbol{A}\boldsymbol{x}+\boldsymbol{B}\boldsymbol{u}$ and
$\boldsymbol{y}=\boldsymbol{C}\boldsymbol{x}$,
the transfer matrix is
$\boldsymbol{G}(s)=\boldsymbol{C}(s\boldsymbol{I}-\boldsymbol{A})^{-1}\boldsymbol{B}$.
Using the modal expansion of the resolvent, one obtains~\cite{chitsongchen_2013_linear}:
\vspace{-0.15cm}
\begin{equation}
\small
(s\boldsymbol{I}-\boldsymbol{A})^{-1}
=
\sum_{i=1}^{l}
\frac{\boldsymbol{\phi}_i \boldsymbol{\psi}_i^{\top}}{s-\lambda_i}, \label{eig_decompose}
\end{equation}
where $l$ is the total number of system eigenvalues, $\lambda_i$ represents the $i$-th eigenvalue (natural mode), while $\boldsymbol{\phi}_i$ and $\boldsymbol{\psi}_i$ denote the corresponding right and left eigenvectors, respectively. 
By substituting \eqref{eig_decompose} into the transfer function definition, $\boldsymbol{G}(s)$ can be rewritten in its modal expansion form:
\vspace{-0.1cm}
\begin{equation}
\small
\boldsymbol{G}(s)
=
\sum_{i=1}^{l}
\frac{\boldsymbol{C}\boldsymbol{\phi}_i \boldsymbol{\psi}_i^{\top}\boldsymbol{B}}{s-\lambda_i}.
\end{equation}
Defining the output and input participation vectors as
$\boldsymbol{a}_i=\boldsymbol{C}\boldsymbol{\phi}_i$
and
$\boldsymbol{b}_i=\boldsymbol{\psi}_i^{\top}\boldsymbol{B}$,
the transfer matrix becomes
\vspace{-0.1cm}
\begin{equation} 
\small
\label{eq.modal_expansion_final}
\boldsymbol{G}(s)
=
\sum_{i=1}^{l}
\frac{\boldsymbol{a}_i \boldsymbol{b}_i}{s-\lambda_i}.
\end{equation}
   \vspace{-0.8cm}
\subsection{Single-Mode Dominance Analysis}
Near a resonance frequency, the response is often dominated by a single mode $\lambda_r$. Under this assumption,
\vspace{-0.1cm}
\begin{equation}
\small 
\label{eq.single_mode_approx}
\boldsymbol{G}(j2 \pi f)
\approx
\frac{\boldsymbol{a}_r \boldsymbol{b}_r}{j2 \pi f-\lambda_r} = \alpha_r \boldsymbol{a}_r \boldsymbol{b}_r,
\end{equation}
where the scalar frequency coefficient $\alpha_r$ is defined as:
\vspace{-0.1cm}
\begin{equation}
\small 
\alpha_r = \frac{1}{j 2 \pi f - \lambda_r}.
\end{equation}

To compute the sensitivity indicators, we evaluate the maximum singular value ($\sigma^{\max}$) and the corresponding primary left singular vector ($\boldsymbol{u}_1$) of the rank-one matrix approximation in \eqref{eq.single_mode_approx}. By definition, $\sigma^{\max}$ equals the square root of the largest eigenvalue ($\lambda_{\max}$) of the matrix product $\boldsymbol{G}(j2 \pi f)\boldsymbol{G}(j2 \pi f)^{\mathsf{H}}$:
\vspace{-0.1cm}
\begin{equation}
\small
\begin{aligned}
\boldsymbol{G}(j2 \pi f)\boldsymbol{G}(j2 \pi f)^{\mathsf{H}} 
&= 
\left( \alpha_r \boldsymbol{a}_r \boldsymbol{b}_r \right) \left( \alpha_r^* \boldsymbol{b}_r^{\top} \boldsymbol{a}_r^{\mathsf{H}} \right) \\
&= 
|\alpha_r|^2 (\boldsymbol{b}_r \boldsymbol{b}_r^{\mathsf{H}}) \boldsymbol{a}_r \boldsymbol{a}_r^{\mathsf{H}}
\end{aligned}
\end{equation}
Since the matrix is rank one, its sole non-zero eigenvalue corresponds to its matrix trace, leading directly to:
\vspace{-0.1cm}
\begin{equation}
\small 
\lambda_{\max}\left(\boldsymbol{G}\boldsymbol{G}^{\mathsf{H}}\right) = |\alpha_r|^2 \left\|\boldsymbol{a}_r\right\|^2 \left\|\boldsymbol{b}_r\right\|^2.
\end{equation}
Taking the square root gives the maximum potential amplification:
\vspace{-0.1cm}
\begin{equation} 
\small
\label{eq.sigma_max_derived}
\sigma^{\max} = \sqrt{\lambda_{\max}} = |\alpha_r| \, \left\|\boldsymbol{a}_r\right\| \left\|\boldsymbol{b}_r\right\| = \frac{\left\|\boldsymbol{a}_r\right\| \left\|\boldsymbol{b}_r\right\|}{|j 2 \pi f-\lambda_r|}.
\end{equation}

The dominant left singular vector is $\boldsymbol{u}_1$, which aligns perfectly with the normalized range space of the matrix, as dictated by the output participation vector:
\vspace{-0.1cm}
%\begin{equation} \label{eq.u1_derived}
$\boldsymbol{u}_1 = \boldsymbol{a}_r/\left\|\boldsymbol{a}_r\right\|$.
%\end{equation}

The local sensitivity indices proposed in this work are fundamentally formulated on the directional amplification product $\left|\boldsymbol{u}_1\right| \sigma^{\max}$. Substituting $\boldsymbol{u}_1$ into \eqref{eq.sigma_max_derived} into this product yields:
\vspace{-0.1cm}
\begin{equation} 
\small
\label{eq.cancellation_step}
\left|\boldsymbol{u}_1\right| \sigma^{\max} = \left| \frac{\boldsymbol{a}_r}{\left\|\boldsymbol{a}_r\right\|} \right| \frac{\left\|\boldsymbol{a}_r\right\| \left\|\boldsymbol{b}_r\right\|}{|j2\pi f-\lambda_r|}.
\end{equation}
In \eqref{eq.cancellation_step}, the output norm term $\left\|\boldsymbol{a}_r\right\|$ directly cancels out. Furthermore, since the input vector $\boldsymbol{B}$ isolates a single scalar injection at a specific bus $n$, the input participation vector reduces to a scalar quantity, meaning its vector norm matches its absolute scalar value ($\left\|\boldsymbol{b}_r\right\| = \left|\boldsymbol{b}_r\right|$). 
As a result of these algebraic simplifications, the $\text{FSI}$ simplifies elegantly to:
\begin{equation}
\small
\text{FSI}(j2 \pi f) = \frac{\left|\boldsymbol{a}_r\right| \left|\boldsymbol{b}_r\right|}{|j2 \pi f-\lambda_r|}.
\end{equation}
Finally, substituting
$\boldsymbol{a}_r=\boldsymbol{C}\boldsymbol{\phi}_r$
and
$\boldsymbol{b}_r=\boldsymbol{\psi}_r^{\top}\boldsymbol{B}$
gives
\begin{equation}
\small
\text{FSI}(j2\pi f)
=
\frac{
|\boldsymbol{C}\boldsymbol{\phi}_r|
|\boldsymbol{\psi}_r^{\top}\boldsymbol{B}|
}
{|j2\pi f-\lambda_r|}.
\end{equation}
\vspace{-0.95cm}
\subsection{Multi-Mode Interaction Analysis}

When multiple modes contribute within the frequency range of interest, the single-mode approximation no longer holds and
\vspace{-0.1cm}
\begin{equation}
\small
\label{eq_multi_final}
\boldsymbol{G}(j2\pi f)
=
\sum_{i=1}^{l}
\frac{\boldsymbol{a}_i\boldsymbol{b}_i}
{j2\pi f-\lambda_i}
=
\boldsymbol{M}_1+\boldsymbol{M}_2+\cdots+\boldsymbol{M}_l,
\end{equation}
where $\boldsymbol{M}_i=\frac{\boldsymbol{a}_i\boldsymbol{b}_i}{j2\pi f\lambda_i}$.
In this case, the dominant singular vectors are determined by the combined contribution of all modal terms rather than a single eigenmode. Consequently, high FSI/VSI values may result from modal interaction in addition to resonance with an individual natural mode.
%This demonstrates why a purely modal approach falls short, and underscores why the proposed SVD-based \ac{fsi}/\ac{vsi} framework is necessary to capture true system vulnerability in multi-modal environments.

% \section*{Acknowledgment}


\vspace{-0.1cm}
\ifCLASSOPTIONcaptionsoff
  \newpage
\fi

\bibliographystyle{IEEEtran}
\bibliography{ref_short.bib}

\end{document}